\documentclass{ieeeaccess}
\usepackage{cite}
\usepackage{amsmath,amssymb,amsfonts}

\usepackage{graphicx}
\graphicspath{{figures/}}

\usepackage{textcomp}
\usepackage{soul}
\usepackage{multirow}
\usepackage{dblfloatfix}

\soulregister\ref7
\soulregister\cite7

\usepackage[caption=false,font=footnotesize]{subfig}
\usepackage{algorithm}%
\usepackage{algpseudocode}%
\usepackage{booktabs}
\usepackage{hyperref}
\usepackage{array}

\usepackage{soul}
\definecolor{softyellow}{RGB}{255,245,180}  
\definecolor{softgreen}{RGB}{198,239,206}  
\sethlcolor{softyellow}

\usepackage{bm}
\makeatletter
\AtBeginDocument{\DeclareMathVersion{bold}
\SetSymbolFont{operators}{bold}{T1}{times}{b}{n}
\SetSymbolFont{NewLetters}{bold}{T1}{times}{b}{it}
\SetMathAlphabet{\mathrm}{bold}{T1}{times}{b}{n}
\SetMathAlphabet{\mathit}{bold}{T1}{times}{b}{it}
\SetMathAlphabet{\mathbf}{bold}{T1}{times}{b}{n}
\SetMathAlphabet{\mathtt}{bold}{OT1}{pcr}{b}{n}
\SetSymbolFont{symbols}{bold}{OMS}{cmsy}{b}{n}
\renewcommand\boldmath{\@nomath\boldmath\mathversion{bold}}}
\makeatother

\def\BibTeX{{\rm B\kern-.05em{\sc i\kern-.025em b}\kern-.08em
    T\kern-.1667em\lower.7ex\hbox{E}\kern-.125emX}}

\begin{document}

\title{Quantum Annealing for Realistic Traffic Flow Optimization: Clustering and Data-Driven QUBO}
\author{\uppercase{R. Rusnáková}\authorrefmark{1}, 
\uppercase{M. Chovanec}\authorrefmark{1}, 
and \uppercase{J. Gazda}\authorrefmark{1}
}

\address[1]{Department of Computers and Informatics, Faculty of Electrical Engineering and Informatics,  Technical University of Košice, 04200 Košice, Slovakia}
\tfootnote{This work was supported by The Slovak Research and Development Agency project no. APVV-23-0512 and the Slovak Academy of Sciences project no. VEGA 1/0685/23.}


\corresp{Corresponding author: R. Rusnáková (e-mail: renata.rusnakova@tuke.sk).}

\begin{abstract}
This article presents a scalable, data-driven formulation of city-wide Traffic Flow Optimization as a Quadratic Unconstrained Binary Optimization problem and evaluates its performance using quantum annealing and classical solvers on realistic urban networks. The framework builds a time-resolved congestion model from simulated mobility data by sampling vehicle trajectories at fixed intervals, identifying leader–follower interactions on shared road segments. In addition to congestion, the model incorporates route-duration penalties and an analytically derived penalty parameter that enforces one-hot route selection, ensuring feasible assignments while balancing network-wide congestion reduction and individual travel times.
To mitigate the combinatorial growth of interactions in large-scale instances, the approach employs Leiden clustering to partition vehicles into dense communities that can be optimized independently. The resulting subproblems are solved using D-Wave's LeapHybridBQMSampler, exact mixed-integer programming via Gurobi, and several classical metaheuristics and are evaluated on multiple city maps with up to 25,000 vehicles. Across large scenarios, the hybrid quantum-classical approach consistently produces feasible solutions within approximately 1\% of Gurobi's objective values, while maintaining stable runtimes. Both methods outperform shortest-route baselines, achieving reductions in the proposed congestion-cost objective of up to 24.4\% for the hybrid solver and 29.4\% for Gurobi.
Finally, the study highlights the critical role of the underlying city map, showing that network structure directly influences interaction density, problem formulation, and the efficiency of embedding and solving on current quantum annealing hardware.
\end{abstract}


\titlepgskip=-21pt

\maketitle

\section{Introduction}
\label{sec:introduction}
\PARstart{O}{ptimization} problems in urban traffic management hold significant practical and economic importance and have received extensive attention in academic and industrial research. On one hand, progress in generating and collecting high-resolution mobility data enables models that capture fine-grained spatiotemporal interactions between vehicles, providing a more realistic representation of traffic conditions.
At the same time, there is growing interest in quantum and quantum-inspired optimization methods across a range of domains, including logistics, finance, and machine learning~\cite{Li2020, Ciacco2025, Volpe2025, Gharehchopogh2023}. Motivated by this trend, the present work focuses on a concrete application of quantum annealing in urban traffic optimization.

We introduce the \textbf{Traffic Flow Optimization (TFO)}, which minimizes congestion modeled as spatiotemporal conflicts between vehicles sharing road segments in the same direction and time, while accounting for route duration. The problem is formulated as a \textbf{Quadratic Unconstrained Binary Optimization (QUBO),} with quadratic terms penalizing conflicting routes, linear terms for travel-time costs, and an analytically derived penalty parameter $\lambda$ that enforces one-hot route assignments without manual tuning~\cite{Traag2019}. For scalability, vehicles are clustered into subproblems using Leiden clustering. Data from OpenStreetMap (OSM) are used for city map extraction, vehicle origins and destinations are generated randomly or around attraction points, and several alternative routes per vehicle are provided via the Valhalla routing engine~\cite{valhalla}. Routes are sampled at fixed-time intervals, enabling congestion computation from leader–follower interactions. QUBOs are solved using D-Wave's quantum annealers, classical metaheuristics (Simulated Annealing, Tabu Search), and exact solvers (CBC, Gurobi), with performance benchmarked against random and shortest-route baselines.

While the present work builds on earlier efforts in quantum traffic optimization, it extends them through more realistic modeling, reproducible workflows, and improved scalability. In particular, the study by Neukart et al.~\cite{Neukart2017} provided an early formulation of TFO using quantum annealing and serves as a reference point for the approach presented in this study.

In their work, Neukart et al.~\cite{Neukart2017} demonstrated the use of a D-Wave's quantum annealer for TFO, and based on this study, Volkswagen demonstrated one of the first real-world applications of quantum annealing in traffic flow optimization with the \textit{Quantum Shuttle} project at the 2019 Web Summit in Lisbon. Using a D-Wave's quantum annealer connected through a cloud-based quantum web service, shuttle buses were dynamically routed based on live traffic conditions. Over four days, the fleet of nine buses completed more than 160 trips, solving over 1,200 optimization tasks~\cite{Yarkoni2020}.
Neukart et al.'s study relies on a static sub-map of Beijing with 418 cars, simplified routing at intersections via the Dijkstra algorithm, and problem instances limited to 1,248 variables solved via \textit{qbsolv}. Random assignment served as the baseline. Our framework extends this by using full OSM-derived city geometries, generating multiple route alternatives per vehicle, computing congestion with respect to travel time from spatiotemporal overlaps, and providing broader benchmarking.

Similarly, Villanueva et al.\cite{Villanueva2025} formulated the problem of minimizing traffic congestion in terms of QUBO and applied the Quantum Approximate Optimization Algorithm (QAOA) to solve it. The limitations of current gate-based quantum hardware restricted their approach to small, 23-vehicle simulated instances. They introduced a novel heuristic variant of QAOA (noise-resilient QAOA), but did not compare its effectiveness and solution quality with classical methods. This study overcomes the scalability limitations by employing a data-driven TFO formulation that leverages Leiden clustering to partition large networks into manageable subproblems.

More recently, Salloum et al.~\cite{Salloum2025} proposed an iterative QUBO-based method for large traffic scenarios, where the problem size is reduced by solving smaller QUBO instances in iterations.
A key difference between our work and the approach of Salloum et al. lies in the handling of route alternatives. While their iterative method generates routes dynamically to resolve conflicts, our framework precomputes all alternatives upfront, thereby eliminating the overhead of repeated route generation.

Unlike prior works restricted to fixed networks~\cite{Mohammed2025, Neukart2017, Yarkoni2020, Villanueva2025, Salloum2025}, our framework generates instances for any city or subregion via center–radius selection, while attraction points support scenarios like events or evacuations, ensuring broad applicability without workflow changes.

Beyond traffic flow optimization, QUBO-based approaches have been applied to broader logistics and scheduling domains. Leib et al.~\cite{Leib2023} addressed a robotic scheduling problem in a laboratory setting, modeling it through three approaches: a QUBO formulation for quantum and quantum-inspired hardware (D-Wave, Fujitsu's Digital Annealer, and hybrid variants) and two mixed-integer programming (MIP) models solved via Gurobi. Their benchmarking showed that digital and hybrid quantum-classical methods remain competitive with classical MIP solvers, despite current quantum hardware constraints. While they used a similar QUBO formulation and solver set, their research focuses on a different optimization problem within the broader transport domain.

Quantum annealing has also been applied to the Vehicle Routing Problem (VRP)~\cite{Tambunan2022, Azad2023}, and similar QUBO-based approaches have been used for assigning ride-pooling services~\cite{Cattelan2024}. These studies, however, do not incorporate time-resolved congestion. The classical Traffic Assignment Problem (TAP) has also been formulated as a QUBO~\cite{Chitty2024}. The proposed TFO is a special, discrete variant of TAP, which allows us to benchmark our approach against existing TAP-based studies and reinforce the validity of our findings.

Congestion reduction is crucial in applications such as commercial navigation platforms, including Google Maps, Waze, and Apple Maps. Google Maps combines live GPS data from millions of users with historical patterns and AI-based traffic prediction, integrating initiatives like Project Green Light for signal optimization~\cite{Wheeler2023, Lau2021}. Waze relies on crowd-sourced reports for dynamic rerouting that minimizes individual travel times~\cite{AminNaseri2022}, while Apple Maps prioritizes privacy-preserving data aggregation. Although these systems excel in scalability and responsiveness, they focus on individual user experience over system-wide optimization and do not expose their algorithms for research integration.

Unlike commercial navigation platforms, IBM's work explores how quantum optimization could support large-scale traffic management in the future, rather than providing direct navigation services to end users. Their research investigates quantum algorithms such as the QAOA for routing and congestion problems, with proof-of-concept studies indicating potential system-wide improvements in urban mobility. A notable example is their collaboration with Ford and the University of Melbourne, as reported by Villanueva et al.~\cite{Villanueva2025}.

The proposed framework can be integrated with existing navigation systems by operating on precomputed route alternatives and traffic data. It acts as an additional optimization layer that adjusts route assignments based on vehicle interactions, with the aim of improving system-level traffic distribution.

\noindent
\textbf{In summary, key contributions of this article are:}
\begin{enumerate}
    \item A fully data-driven, time-resolved QUBO formulation of TFO that captures spatiotemporal vehicle conflicts together with travel-time penalties.
    \item An analytically derived penalty parameter $\lambda$ that enforces one-hot constraints without the need for manual tuning.
   \item A scalable simulation-to-optimization workflow combining Leiden clustering and congestion modeling,  enabling benchmarking of classical and quantum solvers on instances with up to 25{,}000 vehicles, while accounting for the impact of city map diversity.
    \item Experimental evidence that the D-Wave's hybrid solver with a quantum subroutine produces near-optimal solutions (within $\sim$1\% of Gurobi) with stable solver runtimes across different city network structures, achieving a significant reduction in the proposed congestion-cost objective compared to the shortest-route baseline.\footnote{Congestion reductions refer to the objective defined in this work, combining congestion scores and duration penalties.}
    \item An extensible framework that can operate either as a standalone tool for traffic analysis or as an add-on module for navigation platforms supporting congestion-aware routing and hotspot prediction.
\end{enumerate}

The article is organized as follows. Section~\ref{sec:problem} introduces the problem formulation and the corresponding QUBO model. Section~\ref{sec:simulation} describes the benchmark and simulation setup, including tested instances, performance metrics, and the overall workflow. The effectiveness of the proposed framework and its individual components is evaluated in Section~\ref{sec:results}, where we analyze scalability on large-scale instances and compare solver performance. Finally, Section~\ref{sec:discussion} discusses the results, outlines current limitations, and highlights directions for future work.

\section{Problem formulation}
\label{sec:problem}

The TFO problem studied in this work involves assigning each vehicle exactly one route from a set of precomputed alternatives across an entire city network. The formulation balances two key objectives:

\begin{enumerate}
\item Reducing overall congestion by discouraging multiple vehicles from occupying the same road segment in the same direction at overlapping times;
\item Preserving travel efficiency by penalizing routes much longer than each vehicle's shortest option.
\end{enumerate}

This formulation generalizes classical dynamic TAP and quadratic assignment (QAP) problems, both of which are NP-hard~\cite{Johnson1990, Koch2009}. As the number of vehicles and route alternatives increases, the number of binary variables grows linearly, while pairwise interactions increase quadratically. This leads to combinatorial complexity that quickly exceeds the practical limits of exact solvers, but naturally aligns with the quantum annealer's ability to explore dense binary solution spaces in parallel.

Before we formulate the problem mathematically as QUBO, we first explain how the city map, vehicles, and routes are generated, and how congestion weights and the $\lambda$-penalty parameter are calculated.  \\

\noindent
\textbf{City network.} The city is represented as a directed graph $G = (V, E)$, where $V$ is the set of nodes (road intersections and geometry points) and $E$ is the set of directed edges (road segments between nodes). The network is constructed from OSM. Each edge $e\in E$ is annotated with its geometry (sequence of latitude/longitude coordinates), length, travel time estimate, and allowed travel direction. Depending on the experiment setup, we consider either the entire city or a subset defined by a center coordinates and radius set via parameters. \\

\noindent
\textbf{Vehicles and routes.}
We consider a set of vehicles $C = \{1, \dots, n\}$, each with an origin-destination pair.
For every vehicle $i \in C$, $k$ alternative routes are generated using the Valhalla routing engine, typically with $k \in \{2,3\}$.
These routes may differ in duration, length, or spatial overlap.

Each route $R_{i,a_i}$ for vehicle $i$ and route alternative $a_i$ is represented as a sequence of route points sampled every $\alpha$ seconds.
\begin{equation}
R_{i,a_i} = \{ (p_t, e_t, v_t, d_t) \mid t = 0, \alpha, 2\alpha, \dots, T_{i,a_i} \} \;,
\label{eq:vehicle_route}
\end{equation}
where $p_t$ is the vehicle position at time $t$, $e_t \in E$ the traversed edge, $v_t$ the speed, $d_t$ the travel direction at time $t$ and $T_{i,a_i}$ the total travel duration for the route extracted from Valhalla.\\

\noindent
\textbf{Congestion definition.}
\label{par:congestion}
Congestion is measured by how often vehicles end up in a leader-follower situation on the same road segment at the same time step.
Time is divided into steps of length $\alpha$ seconds, and the entire simulation covers an overall time horizon called the time window $w$.
At each step, we group all vehicles that occupy the same edge $e \in E$ and move in the same direction.

Within this group, we only consider ordered pairs $(i,j)$ where vehicle $i$ is physically ahead of vehicle $j$ along the edge direction, the leader and the follower.\footnote{
The leader--follower ordering is determined by their latitude--longitude positions projected onto a unit vector derived from the edge's cardinal direction. The vehicle with the larger projected value is considered to be further downstream (the leader).}
The distance $d_{i,j}(t,a_i,a_j)$ between vehicles $(i,j)$ at time $t$ under route alternatives $a_i$ and $a_j$ is computed using the Haversine formula ~\cite{Robusto1957}, which calculates great-circle distances from GPS coordinates and therefore accounts for the curvature of the Earth.

To quantify congestion, we assign a score to each leader-follower pair that reflects how close they are relative to their speed as
\begin{equation}
\mathrm{score}_{i,j}(e,t,a_i,a_j) = \alpha \cdot \max\!\left(1 - \frac{d_{i,j}(t,a_i,a_j)}{\gamma \,\bar v_{i,j}(t,a_i,a_j)}, \, 0\right)\;,
\label{eq:score}
\end{equation}

with $\bar v_{i,j}(t,a_i,a_j) = \tfrac{1}{2}(v_i(t,a_i) + v_j(t,a_j))$.
The score is measured in seconds, and the ratio $d_{i,j}/\bar v_{i,j}$ has the interpretation of time headway~\cite{Treiber2000, Brackstone1999}, i.e., the time needed for the following vehicle to reach the current position of the leader under their current average speed. The parameter $\gamma$ acts as a headway-sensitivity parameter that determines how strongly close leader--follower interactions contribute to the congestion score. In particular, interactions with time headway larger than $\gamma$ contribute zero (no congestion interaction is recorded), whereas smaller time headways produce a positive congestion contribution that increases as the vehicles become closer in space and time.
Parameter $\alpha$ (the time step) scales the accumulated interaction duration. An illustrative example is provided in Appendix~\ref{app:gamma}.

Finally, for each edge $e \in E$, vehicle pair $(i,j)$ with $i$ ahead of $j$, and their chosen route alternatives $a_i$ and $a_j$, the congestion entry stored in the database is

\begin{equation}
\mathrm{Cong}(e,i,j,a_i,a_j)
=
\sum_{t \leq \tau_{i,j,a_i,a_j}}
\mathrm{score}_{i,j}(e,t,a_i,a_j)\;,
\label{eq:congestion_entry}
\end{equation}

where $\tau_{i,j,a_i,a_j}=\min(T_{i,a_i},T_{j,a_j},w)$ is the effective observation horizon for the vehicle pair.

It is important to note that the proposed metric is not intended to reproduce the full dynamics captured by macroscopic traffic-flow models. Effects such as queue formation, spillback propagation, signal-induced delays, and explicit road-capacity constraints are therefore not modeled explicitly. Instead, the congestion score captures the potential for congestion interactions between vehicles based on their proximity and motion along shared route segments.
Consequently, the metric should not be interpreted as a direct estimate of standard traffic-engineering quantities such as travel-time delay, queue length, or throughput. Rather, it serves as an \textit{interaction-based proxy}: higher vehicle interaction density is associated with reduced headways and an increased likelihood of delays, whereas lower interaction density generally reflects smoother flow conditions.\\

\noindent
\textbf{Congestion weights.}
From the per-edge congestion entries in Eq.~\ref{eq:congestion_entry}, we derive pairwise congestion weights between vehicles and their route alternatives.
Specifically, for each pair of vehicles $i,j \in C$, routes $a_i,a_j \in \{1,\dots,k\}$, and over the entire time window $w$, we define the aggregated weight as
\begin{equation}
w_{i,j,a_i,a_j} = \sum_{e \in E} \mathrm{Cong}(e,i,j,a_i,a_j) \;,
\label{eq:pairwise_weight}
\end{equation}
where the sum includes all edges $e$ where vehicles $i$ and $j$ traverse the same segment in the same direction with $i$ ahead of $j$.\footnote{Because weights are calculated from congestion score, they are also expressed in seconds.}
Note that the congestion weights are inherently directional, as they are computed from leader–follower interactions defined at the edge and time-step level. The weight $w_{i,j,a_i,a_j}$ aggregates contributions over all instances where vehicle $i$ with selected route $a_i$ acts as the leader and vehicle $j$ with selected route $a_j$ as the follower, and in general $w_{i,j,a_i,a_j} \neq w_{j,i,a_j,a_i}$.
Collecting these values for all vehicle pairs and route alternatives yields a 4-dimensional congestion weight tensor
\begin{equation}
W = \{\, w_{i,j,a_i,a_j} \;\mid\; i,j \in C, \; a_i,a_j \in \{1,\dots,k\} \,\} \;.
\label{eq:weight_tensor}
\end{equation}
Invalid vehicle-route combinations (i.e., cases where a vehicle has fewer than $k$ alternatives) receive zero weight. This tensor $W$ encodes the core congestion interaction costs for the QUBO formulation.\\

\noindent
\textbf{Duration penalty.}
In addition to congestion, we incorporate a penalty term to discourage vehicles from being assigned route alternatives that are significantly longer than their fastest available option.\footnote{The penalty can be based on duration or distance, which is decided by setting an input parameter for all computations. In our simulations, we used the shortest duration as the fastest option.}
For each vehicle $i \in C$ and route $a_i \in \{1,\dots,k\}$ with duration $\mathrm{dur}_{i,a_i}$, we define the penalty as
\begin{equation}
\pi_{i,a_i} = \mathrm{dur}_{i,a_i} - \min_{b_i \in \{1,\dots,k\}} \mathrm{dur}_{i,b_i} \;,
\label{eq:duration_penalty}
\end{equation}
which measures the additional travel time compared to the shortest route of vehicle $i$ (in seconds).
This ensures that selecting routes much longer than the best available option incurs an extra cost.
The reasoning is that while congestion avoidance is important, the solution should not force drivers onto unreasonably long detours. \\

\noindent
\textbf{One-hot constraint and penalty parameter $\lambda$.}
To ensure each vehicle selects exactly one route, we impose the one-hot constraint:
\begin{equation}
\sum_{a_i=1}^k x_{i,a_i} = 1, \qquad \forall i \in C \;,
\label{eq:one_hot}
\end{equation}
where $x_{i,a_i} \in \{0,1\}$ equals 1 if vehicle $i$ selects route $a_i$, and 0 otherwise.

To calculate the penalty parameter $\lambda$, we first compute, for each vehicle $i \in C$ and route $a_i \in \{1,\dots,k\}$, the total directional interaction strength with all other vehicles and routes as 
\begin{equation}
\Lambda_{i,a_i} = \sum_{j \in C, j \neq i} \; \sum_{a_j=1}^k \bigl(w_{i,j,a_i,a_j}+w_{j,i,a_j,a_i}\bigr) \;.
\label{eq:lambda_local}
\end{equation}
We then set the global penalty parameter as
\begin{equation}
\lambda = \max_{i \in C,\, a_i \in \{1,\dots,k\}} \bigl( \pi_{i,a_i} + \Lambda_{i,a_i} \bigr) \;.
\label{eq:lambda}
\end{equation}

The penalty parameter is derived analytically rather than tuned empirically, following the Verma--Lewis principle~\cite{Verma2022}. Removing a feasible route assignment for vehicle $i$ on route $a_i$ reduces the objective by at most the duration saving $\pi_{i,a_i}$ plus the full congestion interaction strength $\Lambda_{i,a_i}$. Both terms must therefore appear in the definition of $\lambda$ to guarantee that feasible assignments are always preferred by the solver.
The detailed reasoning and comparison with related approaches are provided in Appendix~\ref{app:lambda}.\\

\noindent
\textbf{QUBO formulation.}\label{par:qubo}
Using binary decision variables $x_{i,a_i} \in \{0,1\}$ for every vehicle $i \in C$ and route alternative $a_i \in \{1,\dots,k\}$,
we separate the QUBO objective into a cost term $q(\mathbf{x})$ and one-hot penalty term $p(\mathbf{x})$.

The cost term encodes pairwise leader--follower congestion interactions (Eq.~\ref{eq:pairwise_weight}) and duration penalties (Eq.~\ref{eq:duration_penalty})
\begin{equation}
\small
q(\mathbf{x}) \;=\;
\sum_{i\neq j} \sum_{a_i=1}^k \sum_{a_j=1}^k w_{i,j,a_i,a_j}\, x_{i,a_i} x_{j,a_j}
\;+\; \sum_{i=1}^{n} \sum_{a_i=1}^{k} \pi_{i,a_i}\, x_{i,a_i}.
\label{eq:qubo_cost}
\end{equation}

The penalty term enforces the one-hot constraint (Eqs.~\ref{eq:one_hot} and \ref{eq:lambda})
\begin{equation}
p(\mathbf{x}) \;=\; \lambda \sum_{i=1}^{n} \Big( 1 - \sum_{a_i=1}^{k} x_{i,a_i} \Big)^2 .
\label{eq:qubo_penalty}
\end{equation}

The full QUBO objective is then given by
\begin{equation}
\mathcal{Q}(\mathbf{x}) \;=\; q(\mathbf{x}) \;+\; p(\mathbf{x}).
\label{eq:qubo_objective}
\end{equation}

The optimization problem is then to find the vehicle-route assignment $\mathbf{x}^\star$, such that the overall objective is minimal
\begin{equation}
\mathbf{x}^\star \;=\; \arg\min_{\mathbf{x}\in\{0,1\}^{n\cdot k}} \; \mathcal{Q}(\mathbf{x}).
\label{eq:qubo_min}
\end{equation}

Because the QUBO is generated directly from the stored routes, congestion weights, and route alternatives, the process is fully reproducible, and the results are easy to interpret and visualize. The constructed QUBO matrix then serves as the direct input to the optimization solvers. This ensures that each solver operates on an identical problem representation, enabling consistent comparison across methods. The detailed algorithm for constructing the QUBO matrix is provided in Appendix~\ref{app:alg_qubo_marix}.

\section{Benchmark and simulation setup}
\label{sec:simulation}
To evaluate the QUBO formulation of the introduced TFO as defined in (Eq.~\ref{eq:qubo_objective}), we designed a simulation workflow that generates reproducible problem instances, applies different solvers, and records detailed performance metrics.

\subsection{Configuration parameters and QUBO matrix}
Table~\ref{tab:params} shows the key configuration parameters, short descriptions, and example values. These parameters control every stage of the simulation workflow.

\begin{table}[h]
\caption{Configuration parameters for simulation generation. Each parameter links directly to the mathematical notation introduced in Section~\ref{sec:problem}.}
\label{tab:params}
\small
\setlength{\tabcolsep}{3pt}
\begin{tabular}{p{25pt}p{145pt}p{65pt}}
\hline
\textbf{Par} & \textbf{Description} & \textbf{Value} \\
\hline
$C_\text{name}$ & City network for the simulation & Košice, Slovakia \\
$C_\text{coord}$ & Center coordinates $(lat, lon)$  & (48.72°, 21.26°) \\
$C_\text{rad}$ & Radius around $C_\text{coord}$  in [km] & 2 \\
$C_\text{att}$ & Point of interest  & (48.71°, 21.25°) \\ 
$n$ & Number of vehicles $|C|$ & 25000 \\
$L_{\min}$ & Minimal origin-destination length in [m] & 600 \\
$L_{\max}$ & Maximal origin-destination length in [m]& 8000 \\ 
$k$ & Route alternatives per vehicle & 2 \\
$\alpha$ & Route sampling interval in [s] & 10 \\
$w$ & Simulation duration in [s] & 600 \\ 
$\gamma$ & Sensitivity in congestion score in [s] & 4.0 \\ 
$\rho$ & Resolution for Leiden clustering & 4.0 \\
$m$ & Minimum vehicles per cluster & 1000 \\
$L$ & Maximum number of clusters solved & 25 \\
\hline
\end{tabular}
\end{table}

\vspace{1em}
\noindent
Each benchmark instance is therefore uniquely defined by:
\begin{itemize}
    \item \textbf{the city network} $G=(V,E)$ extracted from OSM, configured by $C_\text{name}$, $C_\text{coord}$, $C_\text{rad}$. For testing we used Košice, Prague and Cardiff, with a different radius around the city center.

    \item \textbf{the set of vehicles} $C$ and their origin–destination pairs, defined by $n, L_{\min}, L_{\max}$.
    We restricted trips to $L_{\min}=600$ m (avoiding trivial short trips) and $L_{\max}=8000$ m (within realistic intra-city range), and used an optional central attraction point $C_\text{att}$ to simulate directional flows.

    \item $\boldsymbol{k}$ \textbf{alternative routes} per vehicle generated by the Valhalla routing engine,
    sampled using $\alpha$ and $w$.
    We fix $k=2$ to provide sufficient routing diversity, sample trajectories every $\alpha=10$ seconds to capture leader--follower interactions while keeping trajectory processing computationally manageable, and limit the simulation horizon to $w=600$ seconds to balance realism with solver runtime.

    \item \textbf{the congestion weights} $w_{i,j,a_i,a_j}$ and \textbf{penalties} $\pi_{i,a_i}$, computed as in Eqs.~\ref{eq:pairwise_weight}--\ref{eq:duration_penalty}, are controlled by the sensitivity parameter $\gamma$, which we set to $4.0$ (further explanation is provided in Appendix~\ref{app:gamma}).

    \item \textbf{the optional Leiden clustering} of vehicles into smaller dense subgroups, configured by $\rho, m, L$.
    We fixed $\rho=4.0$, which produces small dense clusters~\cite{Traag2019}, while $m$ (minimum cluster size) and $L$ (maximum number of clusters) were varied across experiments to study their effect on solver scalability and solution quality. For more details, see Subsection~\ref{sec:scalability} and Appendix~\ref{app:clustering}.
\end{itemize}

\subsection{Quantum and classical solvers}\label{sec:solvers}
In our benchmark, we solve the generated instances of TFO using a selection of solver types covering quantum annealing, classical metaheuristic, and exact optimization methods described in more detail in Appendix~\ref{app:solvers}. \\

\noindent
\textbf{Quantum Annealing.}
D-Wave's Systems provides cloud-based access to quantum annealing hardware optimized for QUBO problems~\cite{DWaveAdvantage}; access is programmatic via the Python Ocean SDK through the Leap service. For our testing, we employ either the Advantage QPU directly or the hybrid BQM solver (LeapHybridBQMSampler). \\

\noindent
\textbf{Gurobi.}
Gurobi is an industry-grade commercial solver for MILP~\cite{gurobi}.
It employs a linear-programming-based branch-and-bound algorithm with cutting-plane enhancements,
as originally introduced by Land and Doig~\cite{Land2010}. For our simulations, we use Gurobi via the Python \texttt{gurobipy} package distributed on PyPI~\cite{gurobi-pypi}.\\

\noindent
\textbf{CBC.}
The COIN-OR Branch-and-Cut (CBC) solver~\cite{CBC} is an open-source MILP solver built on a branch-and-bound framework enhanced by cutting planes. We accessed CBC through the \texttt{PuLP} Python library~\cite{PuLP} and executed it locally. \\

\noindent
\textbf{Simulated Annealing (SA).}
This method provides a classical stochastic baseline for QUBO solving. Inspired by the physical cooling process of materials, it explores the solution space through iterative local modifications~\cite{Kirkpatrick1983}. We implemented SA using D-Wave's open-source \texttt{neal} library with default parameters~\cite{Neal}.\\

\noindent
\textbf{Tabu Search (Tabu).}
This metaheuristic enhances local search with short-term memory to avoid cycling. It maintains a dynamic "tabu list" of forbidden moves for a specified tenure (a fixed number of iterations)~\cite{Glover1997}. For our experiments, we used D-Wave's \texttt{dwave-tabu} sampler~\cite{DWaveTabu2025,Palubeckis2004} with default parameters.

\subsection{Scalability via clustering and filtering}\label{sec:scalability}

City-scale TFO instances generated by the proposed simulation framework involve thousands of vehicles, each with multiple alternative routes, leading to QUBO formulations with tens of thousands of binary variables and millions of pairwise interaction terms. Such problem sizes can easily exceed the practical limits of both hybrid quantum-classical and classical solvers. To solve the optimization task without discarding the most relevant congestion conflicts, we employ a clustering strategy.

Following Eq.~\ref{eq:pairwise_weight}, we first construct a weighted congestion graph $G_c = (C, E_c)$ where nodes represent vehicles $C = \{1, \dots, n\}$ and weighted edges $E_c$ represent pairwise congestion interactions. Each undirected edge $\{i,j\} \in E_c$ is weighted by the aggregated congestion cost across all alternative routes and both vehicle orderings:
\begin{equation}
w_{i,j}
=
\sum_{a_i=1}^{k}\sum_{a_j=1}^{k}
\left(
w_{i,j,a_i,a_j}
+
w_{j,i,a_j,a_i}
\right).
\label{eq:cluster_weight}
\end{equation}
This projection reduces the original 4D weight tensor from Eq.~\ref{eq:weight_tensor} to a 2D undirected graph structure suitable for the Leiden community detection algorithm, see Algorithm~\ref{alg:clustering}.

For each detected cluster $C_\ell$, the QUBO formulation and the associated penalty parameter $\lambda$ are constructed using only the vehicles within that cluster. Consequently, the one-hot constraint is defined locally for each cluster based on its corresponding set of vehicles, ensuring feasibility within each optimization subproblem. Further details are provided in Appendix~\ref{app:clustering}.

\begin{algorithm}[h]
\caption{Clustering for scalable QUBO construction}
\label{alg:clustering}
\begin{algorithmic}[1]
\State \textbf{Input:} Directional congestion weights $w_{i,j,a_i, a_j}$, resolution $\rho$, minimum cluster size $m$
\State Build congestion graph $G_c = (C,E_c)$ with edge weights $w_{i,j}$ from Eq.~\ref{eq:cluster_weight}
\State Apply Leiden algorithm with resolution $\rho$ to detect communities
\State Merge small clusters  into larger neighbors by maximum inter-cluster weight until $|C_\ell| \geq m$
\State Repeat previous step until $L$ clusters are built
\State \textbf{Output:} $L$ clusters of vehicles: $\{C_1,\dots,C_L\}$
\end{algorithmic}
\end{algorithm}

\subsection{Instances}
\label{sec:instances}

For benchmarking, problem instances of the TFO are grouped by their size, measured in the total number of binary variables $n_\text{var} = n \cdot k,$
with $n$ vehicles and $k$ alternative routes per vehicle.
To cover different solver capacities, we distinguish three categories of instances:
\begin{enumerate}
    \item \textbf{Small-scale instances} ($n_\text{var} \leq 200$), where direct execution on the QPU is feasible. These instances were evaluated also using solvers from Subsection~\ref{sec:solvers}.
    \item \textbf{Medium-scale instances} ($n_\text{var} \leq 1{,}000$), which fit within the limits of all considered solvers without clustering. These were used for systematic benchmarking across quantum annealing hybrid solver (QAHS), and classical approaches.
    \item \textbf{Large-scale instances} ($n_\text{var} \leq 20{,}000$), where solver capacity becomes a limiting factor.\footnote{The maximum number of linear and quadratic coefficients, a bias limit of $200$ million~\cite{dwave-hybrid-limit}, was reached in the TFO instance with $10{,}000$ vehicles. These limits are inherently problem dependent ~\cite{Quinton2025}.}
    In practice, we also generated larger networks with $25{,}000$ vehicles; however, due to the QAHS limits, clustering was applied to reduce the subproblems' sizes to at most $10{,}000$ vehicles.
    To reflect diverse traffic patterns, vehicle origins and destinations were sampled either randomly or directed toward a common attraction point $C_{\text{att}}$, representing high-demand areas such as stadiums or city centers.
\end{enumerate}

This categorization provides a structured way to analyze solver behavior across scales: from direct QPU runs on small networks, through balanced benchmarks at intermediate sizes, to large-scale city scenarios approaching tens of thousands of binary variables and reflecting real-world traffic situations. Solver coverage across these categories is summarized in Table~\ref{tab:solver_coverage}.

\begin{table}[h]
\centering
\normalsize
\caption{Solver coverage across instance size categories. For very large-scale cases (up to $25{,}000$ vehicles), clustering was applied to reduce subproblem size to at most $10{,}000$ vehicles, in line with hybrid solver limits.}
\label{tab:solver_coverage}
\setlength{\tabcolsep}{3pt}
\begin{tabular}{
>{\raggedright\arraybackslash}p{68pt}
>{\centering\arraybackslash}p{20pt}
>{\centering\arraybackslash}p{27pt}
>{\centering\arraybackslash}p{31pt}
>{\centering\arraybackslash}p{16pt}
>{\centering\arraybackslash}p{20pt}
>{\centering\arraybackslash}p{18pt}
}

\hline
\textbf{Instance size} & \textbf{QPU} & \textbf{QAHS} & \textbf{Gurobi} & \textbf{SA} & \textbf{Tabu} & \textbf{CBC} \\
\hline
Small-scale ($n_\text{var} \leq 200$)      & \checkmark & -- & \checkmark & \checkmark & \checkmark & \checkmark \\
Medium-scale ($n_\text{var} \leq 1{,}000$)     & -- & \checkmark & \checkmark & \checkmark & \checkmark & -- \\
Large-scale ($n_\text{var} \leq 20{,}000$) & -- & \checkmark & \checkmark & -- & -- & -- \\
\hline
\end{tabular}
\end{table}

\subsection{Performance metrics}
\label{sec:performance_metrics}

Solver performance is assessed along two aspects: solution quality, expressed by the achieved objective value, and computational effort.
For each solver run, we record three components:
(i) the solver runtime, defined as the wall-clock time required to obtain an optimal or near-optimal solution,
(ii) the preprocessing time needed to construct the problem in a form acceptable to the solver,
and (iii) the objective value returned by the solver.
When the TFO is clustered and only a subset of vehicles is optimized directly, the global objective value is recomputed from the combined assignments of all vehicles.

To provide fair comparisons, computational time limits are aligned across solvers.
SA, Tabu, and  QAHS are configured with iteration or sampling limits that scale with $n_\text{var}$,
while Gurobi and CBC are restricted to the same runtime as the QAHS (resp. QPU).%
\footnote{We also tested Gurobi with extended time limits; however, the solution quality did not improve significantly compared to the restricted runs, more details in Subsection~\ref{sec:large}}.

Feasibility is enforced by the one-hot constraint, Eq.~\ref{eq:one_hot}; infeasible solutions are flagged, but still included in congestion statistics.
For each TFO instance, the following metrics are collected and stored in the database for further analysis and visualization:
\begin{itemize}

    \item \textbf{Congestion cost} evaluated over the entire simulation as a global recomputation of the cost term (Eq.~\ref{eq:qubo_cost}) from the final route assignment $\mathbf{x}$ for all vehicles:
    \begin{equation}
    \text{Cost}(\mathbf{x}) \;=\; q(\mathbf{x}) \;,
    \label{eq:cong_cost}
    \end{equation}
    where $q(\mathbf{x})$ is evaluated using the complete set of selected routes for all vehicles, including interactions between vehicles belonging to different clusters. In other words, clustering is used only during optimization to decompose the problem into tractable subproblems, whereas the reported congestion cost is recomputed afterwards on the full congestion map and therefore includes both intra-cluster and cross-cluster interactions.

    \item $\boldsymbol{\Delta}$ \textbf{Energy} as the relative energy gap, i.e., the normalized difference between the objective values obtained by two solvers. For instance, comparing energy of QAHS against Gurobi is defined as
    \begin{equation}
     \Delta \text{E}  = \Delta \text{Energy}_{\text{QAHS - Gurobi}} =
    \frac{\text{E}_{\text{QAHS}} - \text{E}_{\text{Gurobi}}}{\left|\text{E}_{\text{Gurobi}}\right|} \;,
    \label{eq:energy_gap}
    \end{equation}
    where $\text{E}_{\text{QAHS}}$ and $\text{E}_{\text{Gurobi}}$ denote the final objective values (energies) returned by the respective solvers. The absolute value in the denominator ensures a consistent interpretation even when the energies are negative. A negative value of $\Delta \text{E}$ indicates that QAHS outperforms Gurobi, while a positive value indicates the opposite. Because constant terms of the penalty expansion are omitted and negative diagonal coefficients appear due to the one-hot constraint, the resulting QUBO energies may take negative values.

    \item \textbf{Problem preparation time}, the wall-clock time required to translate the QUBO into a form acceptable by the solver.
    \item \textbf{Solver runtime}, refers only to the time spent inside the optimization
    stage after the model has been constructed. For direct QPU runs, we report the D-Wave's hardware \texttt{qpu\_access\_time}~\cite{dwave-timing}, which includes
    programming, annealing, and readout on the quantum processor. For the
    hybrid QAHS solver, the reported \texttt{run\_time} corresponds to the time
    spent inside the cloud-based hybrid optimization service, which may involve
    both classical methods and QPU calls~\cite{dwave-hybrid}. For classical solvers, the runtime is measured as the wall-clock time of the optimization procedure executed locally.

    \item \textbf{Assignment validity}, the flag indicating if the one-hot constraint was satisfied.
\end{itemize}

\subsection{Overall workflow}\label{sec:workflow}
The end-to-end hybrid quantum-classical simulation and optimization workflow is designed to be fully reproducible and scalable.
Each simulation run proceeds through the following sequence of steps, illustrated in Algorithm~\ref{alg:pipeline}, Figure~\ref{fig:workflow}, and described in detail in Appendix~\ref{app:workflow}.

In summary, our benchmark setup ensures that every solver is tested on the same QUBO instances built from traffic data. The collected metrics give a clear picture of how classical and quantum methods perform on solving TFO.

\begin{algorithm}[t]
\caption{Traffic Flow Optimization Pipeline}
\label{alg:pipeline}
\footnotesize
\begin{algorithmic}[1]
\State \textbf{Input:} $C_{name}$ or $(C_{coord},C_{rad})$, $n$, $C_{att}$, $k$, $\alpha$, $w$, $\gamma$, $\rho$, $L_{\min}$, $L_{\max}$

\State City map generation: $G=(V,E)$ using OpenStreetMap
\State Generate vehicle set $C=\{1,\dots,n\}$ with origin--destination pairs (random or toward $C_{att}$), constrained by path length $L_{\min} \leq L \leq L_{\max}$
\For{each $i \in C$}
    \State Generate $k$ route alternatives using the Valhalla engine
    \State Sample route points $R_{i,a_i}$, $a_i=1,\dots,k$ with respect to $\alpha, w$ - Eq.~\ref{eq:vehicle_route}
\EndFor

\State Compute congestion $\mathrm{Cong}(e,i,j,a_i,a_j)$ from route points - Eqs.~\ref{eq:score}, \ref{eq:congestion_entry}
\State Compute weights $w_{i,j,a_i,a_j}$ and penalties $\pi_i(a)$ using Eqs.~\ref{eq:pairwise_weight}, \ref{eq:weight_tensor}, \ref{eq:duration_penalty}

\If{instance exceeds solver limits ($n > 10,000$)}
    \State Build congestion graph and apply Leiden clustering (Algorithm~\ref{alg:clustering})
    \State Obtain clusters $\{C_1,\dots,C_L\}$
    \For{each cluster $C_\ell \in \{C_1,\dots,C_L\}$}
        \State Compute $\lambda_\ell$ using Eqs.~\ref{eq:lambda_local}, \ref{eq:lambda}
        \State Construct QUBO $\mathbf{Q}_\ell$ using Eqs.~\ref{eq:qubo_cost}, \ref{eq:qubo_penalty}, \ref{eq:qubo_objective}
        \State Solve $Q_\ell(x)$
    \EndFor
\Else
    \State Compute $\lambda$ using Eqs.~\ref{eq:lambda_local}, \ref{eq:lambda}
    \State Construct QUBO $\mathbf{Q}$ using  Eqs.~\ref{eq:qubo_cost}, \ref{eq:qubo_penalty}, \ref{eq:qubo_objective}
    \State Solve $Q(x)$
\EndIf

\State Recompute global cost and compare with baselines using Eq.~\ref{eq:cong_cost}
\end{algorithmic}
\end{algorithm}

\begin{figure}[H]
    \centering
    \includegraphics[width=0.55\linewidth]{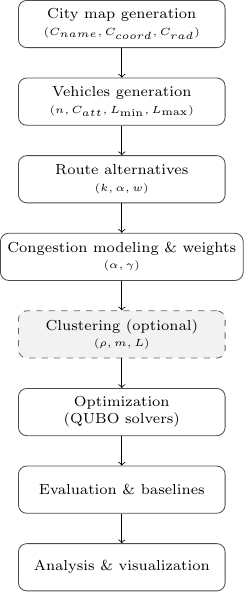}
    \caption{High-level workflow of the traffic optimization. Dashed node indicates an optional
step.}
    \label{fig:workflow}
\end{figure}

\section{Results}\label{sec:results}

Having defined the QUBO formulation and simulation workflow, we now present a detailed evaluation of solvers' performance across problem instances of increasing scale. The results are grouped according to the instance categorization from Subsection~\ref{sec:instances}. All reported congestion reductions refer to the objective defined in this work, combining congestion scores and duration penalties (Eq. ~\ref{eq:qubo_objective}).

\subsection{Small-scale instance analysis}

To initiate the evaluation, we focus on small-scale instances consisting of up to 100 vehicles in city centers of Košice and Prague with 1 km radius and 2 alternative routes. In this setup, all solvers introduced in Section~\ref{sec:solvers} are applied without clustering. These tests allow us to assess solution quality and computational performance on problems where exact solvers like Gurobi and CBC are still tractable, and where quantum annealing can be benchmarked directly against their results.

\begin{figure}[H]
    \centering
    \includegraphics[width=1.0\linewidth]{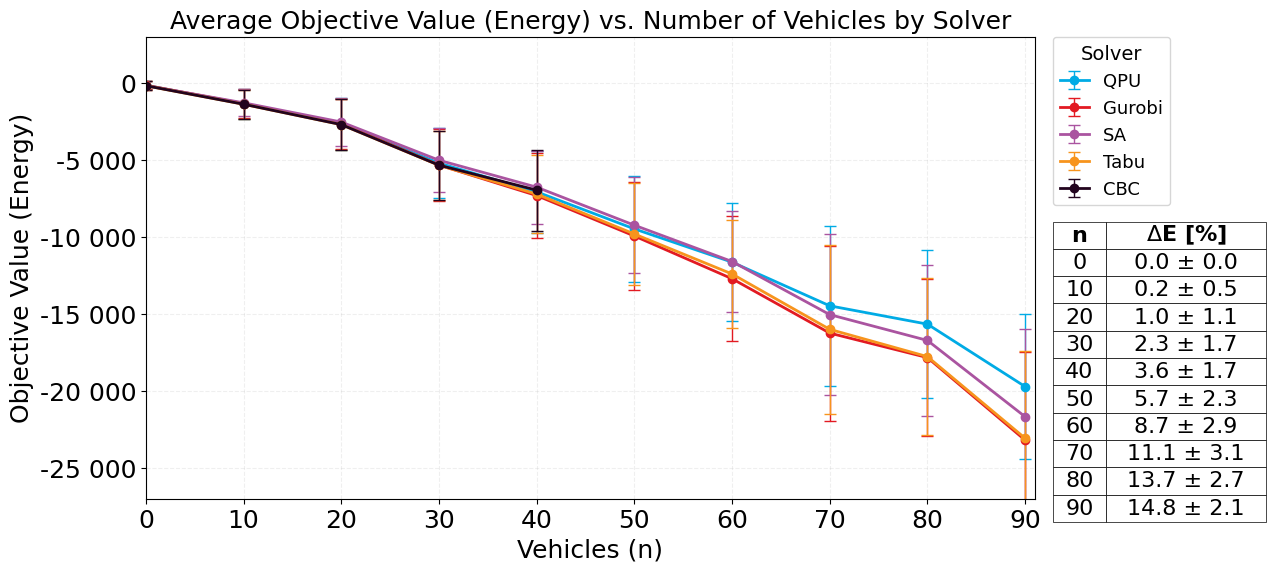}
    \caption{Average objective value (energy) with standard deviation vs. the number of vehicles ($n$) for each solver; evaluated over approximately 100 instances per vehicle count (rounded up).}
    \label{fig:energy_vs_n}
\end{figure}

Figure~\ref{fig:energy_vs_n} presents the average objective value across all simulations, plotted against the number of vehicles (rounded to the nearest ten). As the problem size increases, the objective values decrease due to the growing number of pairwise vehicle interactions encoded in the QUBO formulation. For $n \leq 30$, all solvers perform comparably, with only marginal differences in energy. From $n=40$ onward, Gurobi consistently achieves the lowest objective values, closely followed by Tabu. The QPU, SA, and CBC\footnote{CBC results were included only if the solver status was not ``Not Solved'', indicating that a feasible solution was returned.} remain competitive but begin to diverge as $n$ increases.
The average relative energy gap between QPU and Gurobi remains below 1\% for instances with $n \leq 20$, but increases steadily with problem size, reaching 14.8\% at $n=90$.

\begin{table}[t]
\centering
\caption{Proportion of simulations (\%) in which each solver achieved the lowest or equal objective value; values are reported together with their corresponding standard errors. Entries marked ``--'' indicate configurations that were not evaluated.}
\label{tab:obj_leader}
\scriptsize
\setlength{\tabcolsep}{2pt}
\begin{tabular}
{ >{\centering\arraybackslash}p{35pt} 
>{\centering\arraybackslash}p{40pt} 
>{\centering\arraybackslash}p{41pt} 
>{\centering\arraybackslash}p{38pt} 
>{\centering\arraybackslash}p{38pt} 
>{\centering\arraybackslash}p{38pt} }
\hline
\textbf{Vehicles ($n$)}
& \textbf{QPU [\%]}
& \textbf{Gurobi [\%]}
& \textbf{SA [\%]}
& \textbf{Tabu [\%]}
& \textbf{CBC [\%]} \\
\hline
10  & $100.0\pm0.0$ & $100.0\pm0.0$ & $66.7\pm13.6$ & $83.3\pm10.8$ & $100.0\pm0.0$ \\
20  & $84.2\pm3.2$  & $98.5\pm1.1$  & $20.3\pm3.5$  & $91.0\pm2.5$  & $100.0\pm0.0$ \\
30  & $36.5\pm4.7$  & $98.1\pm1.4$  & $2.9\pm1.6$   & $74.0\pm4.3$  & $98.8\pm1.2$ \\
40  & $3.7\pm1.8$   & $100.0\pm0.0$ & $0.9\pm0.9$   & $46.7\pm4.8$  & $79.6\pm4.2$ \\
50  & $3.0\pm1.7$   & $100.0\pm0.0$  & $0.0\pm0.0$   & $35.0\pm4.8$  & --\\
60  & $2.4\pm1.4$   & $97.6\pm1.4$  & $0.0\pm0.0$   & $23.6\pm3.8$   & --    \\
70  & $0.0\pm0.0$   & $100.0\pm0.0$ & $0.0\pm0.0$   & $11.1\pm3.0$  & --   \\
80  & $1.7\pm1.2$   & $98.3\pm1.2$  & $0.0\pm0.0$   & $9.2\pm2.6$   & --    \\
90  & $2.3\pm2.3$   & $97.7\pm2.3$  & $0.0\pm0.0$   & $0.0\pm0.0$   & --    \\
100 & $0.0\pm0.0$   & $100.0\pm0.0$ & $0.0\pm0.0$   & $14.3\pm13.2$ & --   \\
\hline
\end{tabular}
\end{table}

\begin{table}[h]
\centering
\caption{Proportion of simulations (\%) in which each solver achieved the shortest or equally long runtime; values are reported together with their corresponding standard errors. Entries marked ``--'' indicate configurations that were not evaluated.}
\label{tab:time_leader}
\scriptsize
\setlength{\tabcolsep}{2pt} \begin{tabular}
{ >{\centering\arraybackslash}p{35pt} 
>{\centering\arraybackslash}p{40pt} 
>{\centering\arraybackslash}p{41pt} 
>{\centering\arraybackslash}p{38pt} 
>{\centering\arraybackslash}p{38pt} 
>{\centering\arraybackslash}p{38pt} }
\hline
\textbf{Vehicles ($n$)} &
\textbf{QPU [\%]} &
\textbf{Gurobi [\%]} &
\textbf{SA [\%]} &
\textbf{Tabu [\%]} &
\textbf{CBC [\%]} \\
\hline
10  & $0.0\pm0.0$ & $91.7\pm8.0$ & $8.3\pm8.0$ & $0.0\pm0.0$ & $0.0\pm0.0$ \\
20  & $1.5\pm1.1$ & $98.5\pm1.1$ & $0.0\pm0.0$ & $0.0\pm0.0$ & $0.0\pm0.0$ \\
30  & $1.9\pm1.3$ & $98.1\pm1.3$ & $0.0\pm0.0$ & $0.0\pm0.0$ & $0.0\pm0.0$ \\
40  & $0.0\pm0.0$ & $100.0\pm0.0$ & $0.0\pm0.0$ & $0.0\pm0.0$ & $0.0\pm0.0$ \\
50  & $1.0\pm1.0$ & $99.0\pm1.0$ & $0.0\pm0.0$ & $0.0\pm0.0$ &  -- \\
60  & $2.4\pm1.3$ & $96.9\pm1.5$ & $0.0\pm0.0$ & $0.8\pm0.8$ & -- \\
70  & $0.0\pm0.0$ & $100.0\pm0.0$ & $0.0\pm0.0$ & $0.0\pm0.0$ & -- \\
80  & $1.7\pm1.2$ & $98.3\pm1.2$ & $0.0\pm0.0$ & $0.0\pm0.0$ & -- \\
90  & $2.3\pm2.3$ & $97.7\pm2.3$ & $0.0\pm0.0$ & $0.0\pm0.0$ & -- \\
100 & $0.0\pm0.0$ & $100.0\pm0.0$ & $0.0\pm0.0$ & $0.0\pm0.0$ & -- \\
\hline
\end{tabular}
\end{table}

To further evaluate solver performance, Tables~\ref{tab:obj_leader} and~\ref{tab:time_leader} report the proportion of simulations in which each solver achieved either the best (or tied-best) objective value or the shortest (or tied-shortest) runtime, respectively.

The results confirm that Gurobi outperforms the other solvers on small-scale instances, consistently achieving both the lowest objective values and the shortest runtimes, with an average runtime of only $0.005$ seconds per instance. 

The QPU also performs well for simulations with $n \leq 30$, returning feasible solutions in 93\% of cases and often producing objective values close to those obtained by Gurobi. This observation is consistent with previous studies showing that quantum annealers can provide high-quality solutions for small, densely connected QUBOs~\cite{Neukart2017}.

Tabu search also demonstrates competitive performance on smaller instances, providing a practical classical alternative when exact methods are unavailable. In contrast, although SA is computationally efficient, it appears less effective at exploring the complex energy landscape of dense traffic QUBOs without additional parameter tuning. Due to runtime limitations and increasing computational overhead, CBC remains practical only for instances with $n \leq 40$ and is therefore excluded from further evaluation.

This analysis establishes a baseline for subsequent sections, where we examine how solver performance evolves with increasing problem size, particularly as QAHS is applied to medium- and large-scale instances.

\subsection{Medium-scale instance analysis}

We now consider medium-scale instances containing up to $500$ vehicles. These experiments were conducted in larger urban areas (the city centers of Prague and Košice with a radius up to 1–3 km), characterized by higher vehicle densities and more complex routing interactions. To obtain such dense interaction patterns, we first simulated $n = 5{,}000$ vehicles in each area and then applied the Leiden clustering algorithm to extract compact congestion clusters of size $m \leq 500$, since directly simulating only 500 vehicles over the same region did not produce sufficiently rich congestion structures. In this scenario, we replaced the QPU solver with QAHS and benchmarked its performance against Gurobi, SA, and Tabu.

\begin{figure}[ht]
    \centering
    \includegraphics[width=1.0\linewidth]{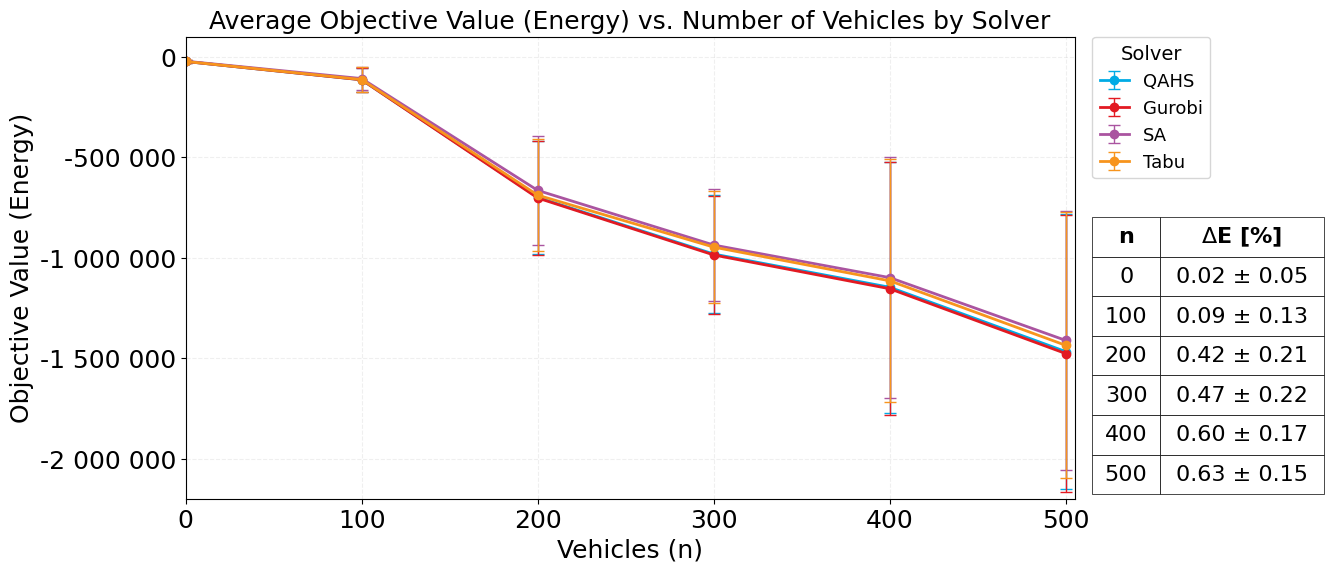}
    \caption{Average objective value (energy) with standard deviation vs. number of vehicles ($n$) for each solver in medium-scale instances; tested for $\approx 50$ instances per $n$ (rounded up).}
    \label{fig:medium_energy_plot}
\end{figure}

As shown in Figure~\ref{fig:medium_energy_plot}, Gurobi consistently achieved the best objective values across all tested instance sizes. QAHS closely followed, with the average relative energy gap increasing gradually from below 0.1\% at $n=100$ to approximately 0.6\% at $n=500$, while producing feasible solutions in 100\% of cases. Although SA and Tabu scaled to larger values of $n$, neither produced competitive objective values. Their deviation from best objective values increased with problem size, and they are therefore excluded from the subsequent large-scale analysis.

Regarding solver runtime, Gurobi remained the fastest solver up to $n \approx 300$, with execution times ranging from milliseconds to a few seconds, depending on problem size and density. QAHS, on the other hand, exhibited a stable runtime of approximately 3 seconds across all instances, which is consistent with the design of D-Wave's hybrid service that enforces a problem-size-dependent minimum time limit and often leads to effectively constant wall-clock times\cite{dwave-hybrid-limit}. This constant-time behavior aligns with existing empirical findings for hybrid solvers on quadratic optimization tasks~\cite{Chitty2024} and will be further discussed in Section~\ref{sec:large}. Tabu and SA showed significantly longer runtimes, especially at higher $n$, with Tabu exceeding 200 seconds per instance in the worst cases; their growing computational cost, combined with declining solution quality, further supports their exclusion from subsequent experiments.

In summary, the medium-scale experiments show that Gurobi and QAHS are the most effective solvers for TFO formulated as a QUBO problem. Gurobi continues to provide optimal solutions efficiently, while QAHS offers a scalable alternative with reliable constraint feasibility, predictable runtimes, and near-optimal solution quality.

In the next section, we extend the evaluation to large-scale city-scale scenarios involving thousands of vehicles, including routing toward both random and predefined destinations, which generate distinct congestion patterns.

\subsection{Large-scale instance analysis}
\label{sec:large}

To assess solver performance at a city scale, we evaluated over 1,500 large-scale instances with up to $n = 25,\!000$ vehicles. These simulations were conducted on city maps of Prague, Košice, and Cardiff, with vehicle routes generated randomly or directed toward attraction-based destinations. At this scale, we focused exclusively on comparing QAHS and Gurobi, as additional classical methods (SA, CBC, and Tabu) exhibited lower solution quality and longer runtimes relative to Gurobi in previous experiments.

\begin{table}[ht]
\caption{Average QAHS performance across large-scale simulations ($n \leq 10,\!000$) rounded to thousands, evaluated on approximately 1,500 TFO instances. The percentage energy gap is reported as mean $\pm$ standard deviation, while validity and dominance rates are reported as proportions of simulations. The last column reports the percentage of simulations in which QAHS achieved an energy lower than or equal to that of Gurobi.}
\label{tab:qa_large_summary}
\footnotesize
\setlength{\tabcolsep}{2pt}
\begin{tabular}{
>{\centering\arraybackslash}p{41pt}
>{\centering\arraybackslash}p{53pt}
>{\centering\arraybackslash}p{45pt}
>{\centering\arraybackslash}p{75pt}
}
\hline
\textbf{Vehicles ($n$)} &
$\boldsymbol{\Delta} \mathbf{E}$ [\%] &
\textbf{Valid [\%]} &
$\mathbf{E}_{\text{QAHS}} \leq \mathbf{E}_{\text{Gurobi}}$ [\%] \\
\hline

500
& $0.66 \pm 0.31$
& $100.00$
& $31.4$
\\

1,000
& $0.63 \pm 0.26$
& $100.00$
& $3.6$
\\

2,000
& $0.76 \pm 0.13$
& $100.00$
& $0.0$
\\

3,000
& $0.80 \pm 0.12$
& $100.00$
& $0.0$
\\

4,000
& $0.90 \pm 0.12$
& $100.00$
& $0.0$
\\

5,000
& $0.90 \pm 0.08$
& $99.98$
& $0.0$
\\

6,000
& $0.94 \pm 0.06$
& $99.99$
& $0.0$
\\

7,000
& $0.98 \pm 0.06$
& $99.98$
& $0.0$
\\

8,000
& $0.91 \pm 0.11$
& $99.98$
& $0.0$
\\

9,000
& $0.88 \pm 0.10$
& $99.99$
& $0.0$
\\

10,000
& $0.99 \pm 0.01$
& $99.98$
& $0.0$
\\

\hline
\end{tabular}
\end{table}

As summarized in Table~\ref{tab:qa_large_summary}, QAHS maintained an
average energy gap of less than $\sim$1\% relative to Gurobi across all
large-scale instances. In a small fraction of cases, primarily for $n \leq 1{,}000$, QAHS achieved energy values lower than or equal to those obtained by Gurobi. This behaviour reflects a scale-dependent transition: at $n = 500$, QAHS matches or outperforms Gurobi in $31.4\%$ of runs, a rate that falls sharply to $3.6\%$ at $n = 1{,}000$ and reaches $0\%$ for all larger instances. Since Gurobi's optimality gap (discussed in more detail at the end of this Section) remains close to zero at these scales, the non-zero QAHS dominance rate at smaller instances suggests that QAHS occasionally reaches solutions of comparable quality.
While the overall gap remained small, isolated outliers appeared in very sparse
QUBOs at higher vehicle counts, where $\Delta \text{E}$ reached up to 1.75\%.
Importantly, feasibility was preserved in nearly all cases, with more
than 99.98\% of solutions satisfying the one-hot constraints.

\begin{figure}[ht]
    \centering
    \includegraphics[width=1.0\linewidth]{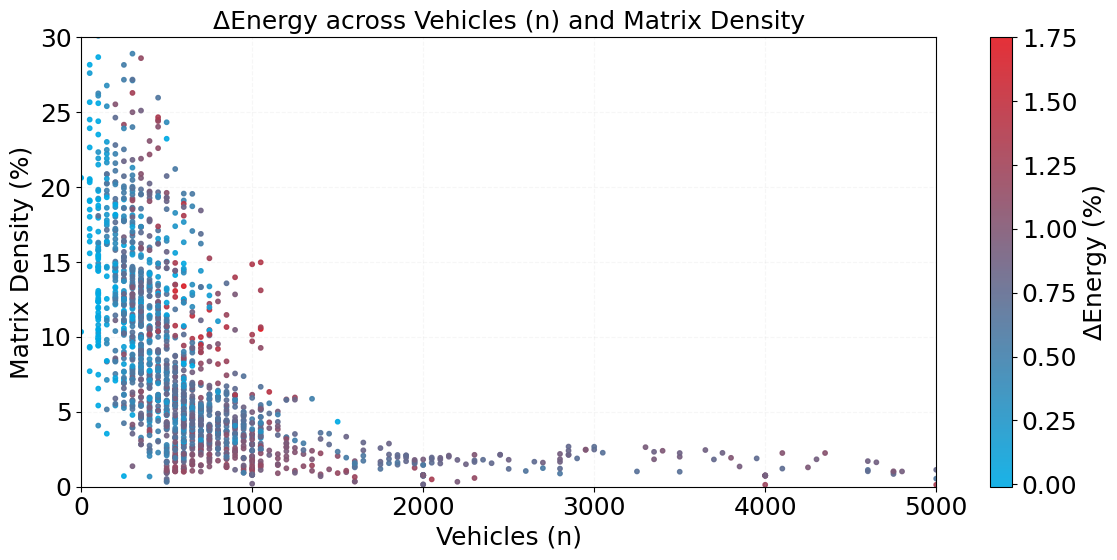}
    \caption{Relationship between solver performance and QUBO density. Each point represents a simulation, with color encoding the QAHS-Gurobi energy gap.}
    \label{fig:delta_energy_density_plot}
\end{figure}

The relationship between QUBO matrix density\footnote{Kim et al.~\cite{Kim2025} define the density of a QUBO matrix as the ratio of the number of quadratic interaction coefficients to the maximum possible interactions $\binom{n_{\text{var}}}{2}$, where $n_{\text{var}}$ is the number of binary variables. We adopt the same calculation.} and solver performance is shown in Figure~\ref{fig:delta_energy_density_plot}. In denser formulations, QAHS closely follows Gurobi, with negligible $\Delta$E, while in sparser formulations the deviation increases, as reflected by higher $\Delta$E values, particularly for larger $n$. This pattern aligns with Kim et al.~\cite{Kim2025}, who observed that quantum annealers are most competitive on dense matrices where classical solvers are challenged by quadratic interaction growth, while sparser instances tend to favor classical methods, explaining the slight increase in $\Delta$E at larger scales.

\begin{figure}[h]
    \centering
    \includegraphics[width=1.0\linewidth]{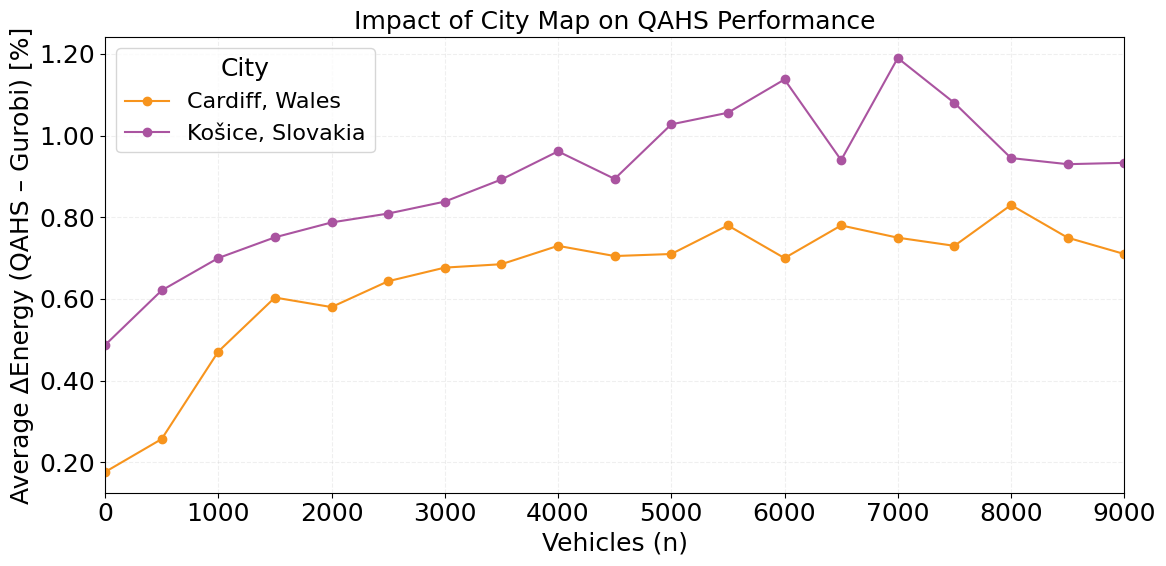}
    \caption{Comparison of solver performance across city networks. Cardiff's more regular topology produces smaller QAHS-Gurobi energy gaps, while Košice's irregular layout results in larger deviations.}
    \label{fig:city_diversity_plot}
\end{figure}

Map diversity also plays a significant role. As shown in Figure~\ref{fig:city_diversity_plot} and Table~\ref{tab:city_delta_energy_compact} (Appendix~\ref{app:tables}), Cardiff achieves the most stable performance, with $\Delta$E remaining below $0.83\%$ even for large $n$, reflecting the advantages of its more regular topology. By contrast, Košice, with its irregular intersections and denser connectivity, exhibits larger energy gaps, indicating that solver performance is more sensitive in irregular, heterogeneous networks. This is an important observation, as the QUBO graph must be embedded onto the solver's physical topology. This embedding step often becomes a bottleneck that can impact solution quality, as also reported in~\cite{Silva2021, Chitty2024}.

\begin{figure}[ht]
    \centering
    \includegraphics[width=1.0\linewidth]{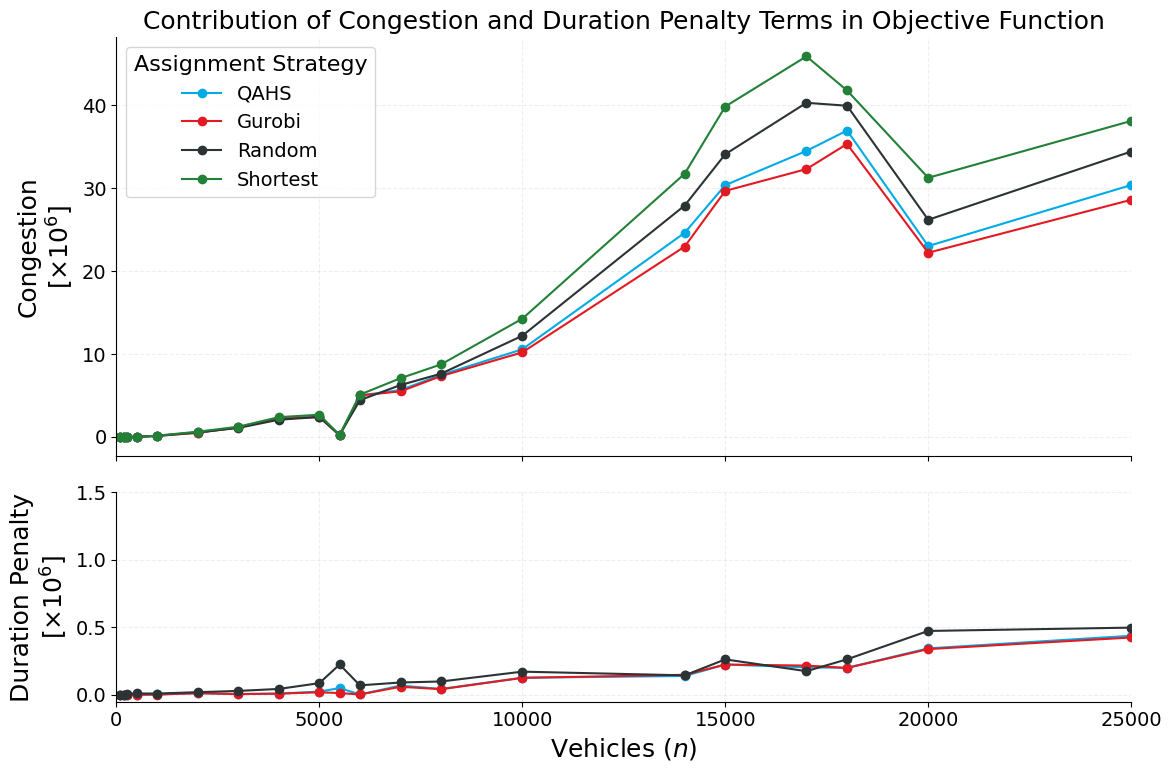}
    \caption{Comparison of congestion and penalty terms of the objective function across assignment strategies (QAHS, Gurobi, random, and shortest-duration baseline). Results are aggregated over large-scale instances (vehicles rounded up).}
    \label{fig:overall_congestion_plot}
\end{figure}

To better assess the system-wide traffic impact, we next compare solvers in terms of congestion cost (Eq.~\ref{eq:cong_cost}), a metric combining route penalties and congestion scores. Figure~\ref{fig:overall_congestion_plot} shows that both QAHS and Gurobi substantially reduce the overall congestion cost relative to the random and shortest-duration baselines, with Gurobi achieving the lowest overall cost.\footnote{The random and shortest-duration strategies are used as baselines because they follow naturally from the proposed QUBO formulation and available route alternatives. Evaluating more advanced congestion-aware heuristics would require additional algorithm design and is therefore left for future work.} The decomposition of the objective function further reveals that the congestion component dominates the total cost and grows approximately as $\mathcal{O}(n^2)$ due to pairwise vehicle interactions, whereas the duration penalty grows only as $\mathcal{O}(n)$ and therefore contributes a comparatively smaller fraction of the objective value as the problem size increases. 
These results provide an additional perspective: assigning every vehicle to its shortest-duration route may appear optimal from an individual perspective, but at the system level, it increases congestion by concentrating traffic on the same high-demand road segments. In contrast, system-oriented optimization allows some vehicles to take slightly longer routes to distribute traffic more evenly across the network, thereby reducing overall congestion.

\begin{table}[ht]
\centering
\caption{Relative congestion-cost improvement achieved by QAHS and Gurobi compared to the shortest-route baseline; tested on the city map of Košice. Values are reported as mean $\pm$ standard deviation.}
\label{tab:shortest_improvement}
\footnotesize
\setlength{\tabcolsep}{2pt}
\begin{tabular}{
>{\centering\arraybackslash}p{42pt}
>{\centering\arraybackslash}p{82pt}
>{\centering\arraybackslash}p{92pt}
}
\hline
\textbf{Vehicles ($n$)} &
\textbf{QAHS vs. Shortest [\%]} &
\textbf{Gurobi vs. Shortest [\%]} \\
\hline
100    & $0.0 \pm 0.0$   & $0.0 \pm 0.0$ \\
200    & $0.0 \pm 0.0$   & $0.0 \pm 0.0$ \\
500    & $8.3 \pm 11.3$  & $15.7 \pm 8.7$ \\
1,000  & $9.4 \pm 9.3$   & $11.6 \pm 12.2$ \\
2,000  & $15.1 \pm 9.5$  & $19.5 \pm 12.5$ \\
3,000  & $7.5 \pm 6.7$   & $7.6 \pm 6.7$ \\
4,000  & $6.0 \pm 8.8$   & $7.3 \pm 11.4$ \\
5,000  & $2.8 \pm 7.4$   & $5.4 \pm 7.5$ \\
6,000  & $1.8 \pm 0.6$   & $1.8 \pm 0.6$ \\
7,000  & $16.3 \pm 12.9$ & $20.7 \pm 10.8$ \\
8,000  & $13.4 \pm 7.9$  & $14.9 \pm 9.2$ \\
10,000 & $22.4 \pm 6.6$  & $24.8 \pm 7.4$ \\
14,000 & $22.0 \pm 0.8$  & $27.2 \pm 1.9$ \\
15,000 & $15.0 \pm 10.6$ & $16.9 \pm 12.0$ \\
17,000 & $24.4 \pm 0.3$  & $29.1 \pm 0.2$ \\
20,000 & $18.8 \pm 9.6$  & $21.4 \pm 11.2$ \\
25,000 & $23.4 \pm 2.5$  & $29.4 \pm 4.4$ \\
\hline
\end{tabular}
\end{table}

From a practical perspective, the value of the presented TFO lies in reducing the proposed congestion-cost objective relative to baseline routing strategies commonly used in existing navigation systems, which typically assign vehicles to the shortest-duration route.
Table~\ref{tab:shortest_improvement} reports the relative reduction in the proposed congestion-cost objective for QAHS and Gurobi compared with the shortest-duration baseline, computed from average costs grouped by the number of vehicles.
Both solvers demonstrate substantial benefits as the number of vehicles increases, with maximum observed average improvements up to 24.4\% (QAHS) and 29.4\% (Gurobi) for large-scale scenarios. For the smallest instances ($n=100$ and $n=200$), no improvement is observed because congestion is low and the shortest-duration routes already provide a good solution, consistent with Figure~\ref{fig:overall_congestion_plot}.

These results highlight two important observations. First, QAHS achieves reductions in the congestion-cost objective comparable to those obtained by a state-of-the-art classical solver, demonstrating competitive performance on large-scale TFO. Second, even moderate percentage improvements can have a meaningful practical impact, since congestion effects tend to increase nonlinearly as the number of vehicles grows.

Next, we discuss the Prague scenario with $n = 15{,}000$ vehicles in two settings: without attraction points and with a single attraction point located at the Main Station. Interactive visualizations of both scenarios are available in the project repository as \texttt{.html} files.\footnote{Congestion maps are publicly available at: \url{https://github.com/rusnakrenata/qa\_mtc/tree/main/maps}.}

In the Prague scenario without attraction point, the optimized solutions distributed traffic more evenly across the network than the random and shortest-duration assignments, which led to more visible congestion hotspots. QAHS and Gurobi achieved the lowest congestion costs, with Gurobi performing slightly better and the relative difference remaining small. As a result, the average delay per vehicle during the 10-minute simulation stayed relatively low at around 3.5\,s.

The attraction-point scenario was more demanding because all vehicles were directed toward a common destination, creating a strong directional flow. In this case, congestion became concentrated on a limited number of access corridors where many routes overlapped. The shortest-duration baseline produced the worst overall result, showing that individually attractive routes can lead to poor network-level outcomes when many vehicles make similar choices. Both QAHS and Gurobi reduced this effect by redistributing part of the flow to alternative route segments, with QAHS again remaining close to Gurobi. Under this stronger demand concentration, the average per-vehicle delay over the 10-minute simulation increased to around 20\,s.

\begin{figure}[ht]
\centering
\includegraphics[width=1.0\linewidth]{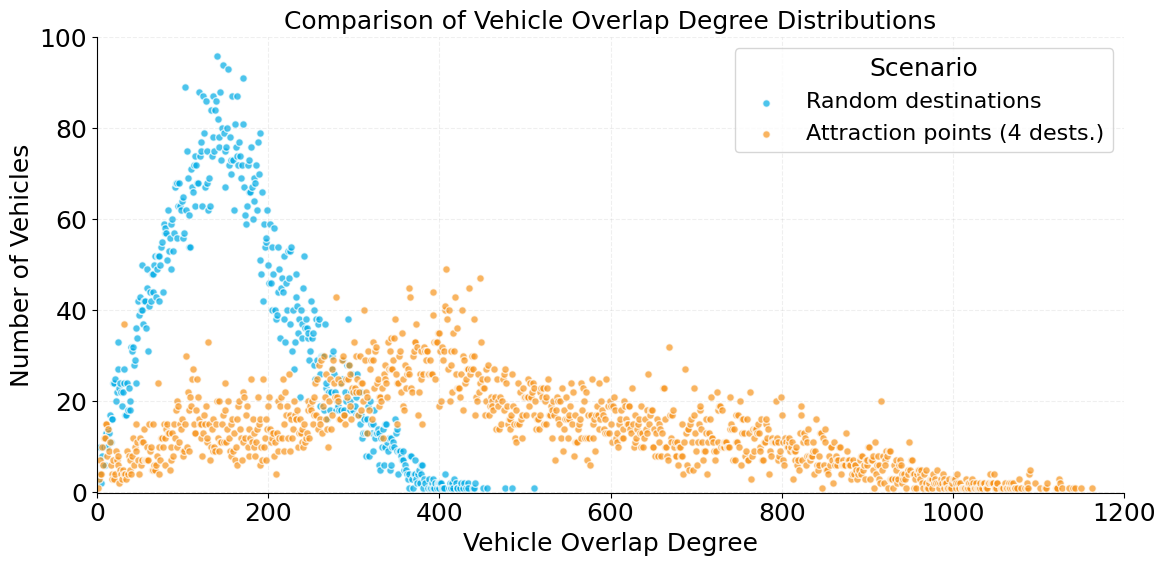}
\caption{Degree distributions of vehicle overlap in the Prague instance for random destinations (blue) and attraction points (orange).}
\label{fig:distribution_plot}
\end{figure}

Figure~\ref{fig:distribution_plot} visually presents how attraction points change the structure of vehicle interactions. For random destinations, the degree distribution of the congestion graph is concentrated at lower overlap degrees, indicating that vehicles interact with a moderate number of others. Most vehicles exhibit overlap degrees below approximately 300, and the number of vehicles decreases rapidly as the degree increases. When attraction points are introduced, the distribution becomes broad and heavy-tailed: a substantial number of vehicles exhibit overlap degrees between 400 and 1,000, while a small group reaches even higher values. In practical terms, this means that several vehicles simultaneously overlap with more than one thousand other vehicles, effectively acting as congestion hubs within the interaction network. This hub-like behavior shows how flows converging to common destinations generate vehicles that accumulate disproportionately high congestion, consistent with empirical observations of non-uniform congestion patterns in real traffic systems~\cite{Wardrop1952, Chen2024}.

\begin{table*}[!t]
\centering
\caption{Summary statistics of the intra-cluster interaction fraction ($\eta$) for different cities and problem sizes. Reported values include the mean ($\eta_{\text{mean}}$), standard deviation ($\sigma_\eta$), and 95\% confidence interval computed across benchmark instances.}
\label{tab:eta_summary}
\setlength{\tabcolsep}{10pt}   
\footnotesize
\begin{tabular}{c|ccc|ccc|ccc}
\hline
&
\multicolumn{3}{c|}{\textbf{Cardiff}}
&
\multicolumn{3}{c|}{\textbf{Ko\v{s}ice}}
&
\multicolumn{3}{c}{\textbf{Prague}}
\\
\cline{2-10}

Vehicles$(n)$
& $\eta_\text{mean}$ & $\sigma_\eta$ & 95\% CI
& $\eta_\text{mean}$ & $\sigma_\eta$ & 95\% CI
& $\eta_\text{mean}$ & $\sigma_\eta$ & 95\% CI
\\
\hline

100
& $0.953$ & $0.000$ & $[0.953, 0.953]$
& $0.842$ & $0.000$ & $[0.842, 0.842]$
& $0.901$ & $0.000$ & $[0.901, 0.901]$
\\

200
& $0.912$ & $0.000$ & $[0.912, 0.912]$
& $0.868$ & $0.028$ & $[0.829, 0.907]$
& $0.857$ & $0.035$ & $[0.808, 0.906]$
\\

500
& $0.875$ & $0.042$ & $[0.827, 0.923]$
& $0.940$ & $0.146$ & $[0.832, 1.000]$
& $0.786$ & $0.044$ & $[0.776, 0.797]$
\\

1,000
& $0.940$ & $0.025$ & $[0.905, 0.975]$
& $0.996$ & $0.004$ & $[0.991, 1.000]$
& $0.769$ & $0.048$ & $[0.722, 0.816]$
\\

2,000
& $0.980$ & $0.018$ & $[0.960, 1.000]$
& $0.994$ & $0.006$ & $[0.986, 1.000]$
& $0.742$ & $0.055$ & $[0.680, 0.804]$
\\

3,000
& $0.840$ & $0.055$ & $[0.764, 0.916]$
& $0.992$ & $0.008$ & $[0.981, 1.000]$
& $0.718$ & $0.065$ & $[0.628, 0.808]$
\\

4,000
& $0.795$ & $0.048$ & $[0.741, 0.849]$
& $0.989$ & $0.011$ & $[0.974, 1.000]$
& $0.704$ & $0.070$ & $[0.625, 0.783]$
\\

5,000
& $0.997$ & $0.003$ & $[0.993, 1.000]$
& $0.824$ & $0.085$ & $[0.750, 0.899]$
& $0.691$ & $0.009$ & $[0.678, 0.704]$
\\

6,000
& $0.658$ & $0.062$ & $[0.597, 0.719]$
& $0.748$ & $0.095$ & $[0.655, 0.841]$
& $0.645$ & $0.072$ & $[0.545, 0.745]$
\\

7,000
& $0.637$ & $0.071$ & $[0.557, 0.717]$
& $0.730$ & $0.110$ & $[0.649, 0.812]$
& $0.595$ & $0.080$ & $[0.504, 0.686]$
\\

8,000
& $0.622$ & $0.058$ & $[0.565, 0.679]$
& $0.695$ & $0.072$ & $[0.637, 0.753]$
& $0.558$ & $0.070$ & $[0.461, 0.655]$
\\

10,000
& $0.619$ & $0.075$ & $[0.553, 0.685]$
& $0.664$ & $0.078$ & $[0.625, 0.704]$
& $0.501$ & $0.000$ & $[0.501, 0.501]$
\\

14,000
& $0.613$ & $0.048$ & $[0.559, 0.667]$
& $0.680$ & $0.068$ & $[0.613, 0.747]$
& $0.643$ & $0.078$ & $[0.555, 0.731]$
\\

15,000
& $0.617$ & $0.070$ & $[0.565, 0.669]$
& $0.705$ & $0.028$ & $[0.674, 0.737]$
& $0.682$ & $0.053$ & $[0.652, 0.713]$
\\

17,000
& $0.611$ & $0.064$ & $[0.548, 0.674]$
& $0.625$ & $0.055$ & $[0.563, 0.687]$
& $0.558$ & $0.065$ & $[0.468, 0.648]$
\\

20,000
& $0.608$ & $0.052$ & $[0.549, 0.667]$
& $0.610$ & $0.062$ & $[0.561, 0.660]$
& $0.496$ & $0.000$ & $[0.496, 0.496]$
\\

25,000
& $0.604$ & $0.067$ & $[0.538, 0.670]$
& $0.633$ & $0.113$ & $[0.549, 0.717]$
& $0.553$ & $0.000$ & $[0.553, 0.553]$
\\

\hline
\end{tabular}
\end{table*}

The runtime analysis reveals important differences between quantum and classical solvers, see Table~\ref{tab:runtime_comparison} (Appendix~\ref{app:tables}). Gurobi consistently outperforms QAHS for smaller problem sizes ($n \leq 4{,}000$), often completing in less than half the time. However, for larger instances, runtimes converge due to a fixed wall-clock limit aligned with QAHS runtime.
To evaluate the impact of the imposed runtime limit on the classical
solver, we analyzed the optimality gap reported by Gurobi across all
experiments. Out of 2,536 runs, 1,200 instances (47.3\%) reached proven
optimality (gap = 0), while 2,475 instances (97.6\%) had an optimality
gap below 1\%. Only 61 instances (2.4\%) exhibited a gap greater than
1\%. These results indicate that the solver typically reaches
near-optimal solutions within the imposed runtime budget.
\footnote{Additional experiments with a 600-second Gurobi runtime limit confirmed that the proposed limit did not affect solution quality, indicating that the QAHS runtime was already sufficient.} Nonetheless, unlike classical MILP solvers such as Gurobi, whose worst-case runtime grows exponentially with problem size (e.g., due to branch-and-bound search, $O(2^n)$)~\cite{Johnson1990}, QAHS maintains more stable empirical scaling by distributing workloads into independent subproblems~\cite{Raymond2023}. As a result, the QAHS solver exhibits highly predictable timing behavior, with runtime increasing almost linearly with problem size, i.e., the number of subproblems generated during decomposition, from just 3 seconds at $n=100$ to approximately 117 seconds at $n=10{,}000$.
Importantly, QPU access time (the actual time spent on the QPU) remains negligible (below 0.1 seconds across all sizes), confirming that the bottleneck lies in orchestration, problem embedding, and postprocessing. These overheads include the programming cycle, thermal stabilization, and communication with classical controllers. Even small problems incur similar initialization costs, which explains the nearly fixed baseline runtime \cite{dwave-timing}.

Regarding the model preparation time, which is executed locally, Gurobi's translation of QUBO to MILP is relatively lightweight, whereas QAHS requires converting the QUBO into a Binary Quadratic Model (BQM). This step introduces additional object-construction and normalization overhead and can take up to twice as long as Gurobi's MILP preparation on our hardware. While faster local machines would reduce both times proportionally, part of this gap is intrinsic to the current QUBO–BQM pipeline and could be narrowed in future versions of the D-Wave's hybrid stack through more efficient preprocessing\cite{dwave-hybrid}.

To quantify how effectively the Leiden decomposition captures the dominant congestion interactions, we define the intra-cluster interaction weight fraction
\begin{equation}
\small
\eta
=
\frac{
\displaystyle\sum_{\ell=1\dots L}\;
\sum_{\substack{(i,j)\in C_\ell}}
w_{i,j}
}{
\displaystyle\sum_{(i,j)}
w_{i,j}
},
\label{eq:eta}
\end{equation}
where $w_{i,j}$ is the total pairwise congestion weight between vehicles $i$ and $j$ summed over all route combinations (Eq.~\ref{eq:cluster_weight}), the numerator aggregates over all pairs $(i,j)$ placed in the same cluster $C_\ell$ (constructed according to Algorithm~\ref{alg:clustering}), and the denominator sums over all pairs in the traffic network. A value of $\eta$ close to $1$ indicates that nearly all congestion interactions fall within jointly optimized subproblems.

Table~\ref{tab:eta_summary} reports the mean intra-cluster interaction fraction $\eta_{\text{mean}}$ across all three city maps and vehicle counts. For instances with $n \leq 5{,}000$, most interaction weight remains concentrated within clusters, indicating that the Leiden algorithm successfully captures the dominant congestion interactions within individual communities.
For Ko\v{s}ice, intra-cluster coverage remains very high over the range $n = 1{,}000$--$4{,}000$, with $\eta_{\text{mean}}$ values between approximately $0.989$ and $0.996$. Cardiff reaches $\eta_{\text{mean}} = 0.997$ at $n = 5{,}000$ and then stabilises around $0.60$--$0.62$ for larger instances, while Prague exhibits a more gradual decline from approximately $0.86$ at small $n$ to $0.50$--$0.55$ for the largest tested sizes.
 
These results should be read alongside the clustering configuration. The minimum cluster size $m$ was varied in the range $100$--$1{,}000$ vehicles and the maximum number of clusters $L$ in the range $10$--$25$, ensuring that the combined capacity $L\times m$ covers the full vehicle set at every tested problem size.
The resolution parameter $\rho = 4$ was held constant across all experiments, as preliminary tests showed that it produces compact communities with strong interactions. 
Most clustering parameter tuning and sensitivity experiments were conducted on the Ko\v{s}ice road network. Consequently, the selected clustering configuration is naturally better aligned with the structural properties of this network, which may contribute to the consistently higher values of $\eta_{\text{mean}}$ observed for Ko\v{s}ice compared to Cardiff and Prague.
Nevertheless, a common trend is visible across all three cities.
As the number of vehicles increases, the number of pairwise congestion interactions grows substantially, making the clustering task increasingly challenging. At the same time, the limits imposed by $m$ and $L$ restrict how finely the congestion graph can be partitioned. Consequently, a larger fraction of interactions remains distributed across cluster boundaries, contributing to the gradual decline in $\eta$.

Notably, although $\eta$ decreases with increasing problem size and falls below $0.9$ for several large-scale configurations, the relative QAHS--Gurobi energy gap remains below $\sim$1\% across all tested configurations (Table~\ref{tab:qa_large_summary}). Since both QAHS and Gurobi optimize the same clustering-based decomposition, the observed gap reflects differences between the optimization methods under the same decomposed formulation rather than the effect of clustering itself. Interactions between vehicles assigned to different clusters are not jointly optimized by either method during the decomposition stage, but are included afterward in the final globally recomputed congestion cost (Eq.~\ref{eq:cong_cost}) used for evaluation.

\section{Discussion}
\label{sec:discussion}

The TFO framework demonstrates that QUBO provides a natural formulation for congestion-aware routing. Its structure aligns closely with D-Wave's quantum annealing architecture, allowing vehicle interactions to be modeled at fine spatial and temporal resolutions and enabling the identification of critical congestion patterns that are often lost in more aggregated models.

This detailed modeling, however, introduces a practical challenge: every pairwise conflict must be stored and included in the QUBO, making matrix construction a potential bottleneck. On a local workstation, preparing the QUBO and computing congestion weights took several minutes, and for the largest instances, this step approached ten minutes. Most of the computational effort is classical and can therefore be offloaded to high-performance computing (HPC) infrastructure, while the optimization stage is handled by specialized solvers.
Such a workflow assumes access to centralized computing resources, for example, in cloud-based systems, where preprocessing can be performed periodically, without the constraints of strict real-time operation. We consider this a current limitation for practical near-real-time deployment. The present evaluation is further limited to a ten-minute static simulation window ($w = 600$\,s) with no vehicle inflow or outflow. The reported per-vehicle delays therefore reflect a short, closed traffic scenario, and how the results extend to longer, multi-hour peak periods with dynamic demand remains an open question left for future work.

The advantage of the proposed approach is most visible in large-scale scenarios with dense vehicle interactions, where the number of pairwise conflicts increases rapidly, and classical heuristics struggle to capture global structure. For smaller or weakly interacting instances, exact classical solvers such as Gurobi remain more suitable.

The baseline routing strategies considered in this study include random
assignment and shortest-duration routing. These methods serve as simple and
reproducible reference strategies for evaluating the effect of the QUBO-based optimization. More advanced classical traffic heuristics, such as greedy congestion-aware route assignment or local-improvement methods, could provide additional points of comparison and may further clarify the relative
advantages of QUBO-based optimization. Although the framework relies on realistic road networks and routing information, validation against real-world mobility traces remains outside the scope of this study.

The key workflow parameters, such as $\alpha$, $\gamma$, and $\rho$, were selected based on commonly adopted settings from related studies and best practices. Preliminary experiments were performed to guide and validate these choices. Nevertheless, a more detailed sensitivity analysis could provide additional insights and is therefore left for future work.

A key strength of the framework is its flexibility. Road closures or blocked junctions can be incorporated immediately by removing the affected segments from the optimization graph. Likewise, traffic incidents reported by external services such as Waze can be used to exclude impacted areas and adjust nearby routes, ensuring that emergency vehicles retain priority access while the remaining traffic is rerouted accordingly.
In addition to reactive control, the framework also supports proactive analysis. Planned events, such as concerts or sporting matches, can be simulated in advance to identify vulnerable corridors and assess mitigation strategies before congestion occurs. In these cases, the hybrid workflow can be executed offline, which we consider one of the main practical advantages of the proposed framework. 

The framework is also particularly relevant in the context of autonomous mobility. In conventional traffic systems, only a subset of drivers follow routing recommendations, which limits the benefits of system-wide optimization. In contrast, when autonomous or centrally coordinated vehicles consistently follow congestion-aware route assignments, the effect of global optimization becomes much stronger, leading to more substantial reductions in congestion and travel time.

Although the current formulation focuses on the reduction of the proposed congestion-cost objective, it can be extended to capture additional aspects of traffic dynamics. Road capacity limits could be introduced through penalties once flows exceed physical thresholds, and delays at intersections could be modeled by incorporating waiting times at traffic signals. While such extensions increase model complexity, they may become increasingly practical as quantum hardware matures and solver connectivity improves, enabling larger and more detailed formulations to be addressed effectively.

\section*{Conclusion}
\label{sec:conclusion}

Our results show that the D-Wave's hybrid solver with a quantum subroutine produces solutions that are consistently close to those obtained by the state-of-the-art classical solver Gurobi, typically within $\sim$1\% of Gurobi's objective values. While Gurobi systematically achieves the best solutions, the hybrid solver exhibits stable runtimes (with QPU access time below 0.1\,s across all problem sizes) and maintains a high rate of feasible assignments across large-scale instances. These findings indicate that hybrid quantum-classical approaches can serve as a viable complementary method for complex optimization problems such as large-scale traffic flow optimization. In addition, compared to the shortest-route baseline, the proposed optimization framework reduces congestion by up to 24.4\% for the hybrid quantum-classical solver, demonstrating a clear benefit for system-wide mobility management.

 Unlike commercial platforms that primarily optimize individual travel times, the presented QUBO formulation balances driver efficiency with the reduction of the proposed congestion-cost objective. This positions the framework as a promising addition to future ITS, where integration with live mobility data could help anticipate hotspots, smooth traffic flow, and improve the use of road networks.

Building on prior works that addressed small test cases or iterative QUBO refinements, our study demonstrates that hybrid quantum-classical optimization (leveraging the LeapHybridBQMSampler, where the QPU serves as a subroutine), combined with classical problem decomposition (e.g., using Leiden clustering), can scale to realistic city networks with up to 25,000 vehicles. Moreover, the provided experiments highlight sensitivity to network topology, with Cardiff showing a gap of 0.83\%, underlining the importance of map diversity in performance evaluation.

Looking ahead, ongoing advances in quantum hardware, solver connectivity, and embedding techniques are expected to further increase the size and quality of problems that can be addressed directly on quantum devices. At the same time, our results highlight the current limitations of D-Wave's system in the context of traffic flow optimization, particularly with respect to achievable problem size and solution quality. These limitations are strongly problem and instance dependent, as different city topologies and modeling choices place different demands on the hardware.

\appendices

\section{Sensitivity parameter}\label{app:gamma}
In the congestion score (Eq.~\ref{eq:score}), the parameter $\gamma$ controls the sensitivity of the model to vehicle spacing. It scales the ratio $\tfrac{d}{\bar v}$, which represents the time headway, a standard measure used in traffic safety and leader–follower dynamics.

By setting $\gamma \in [2,4]$, we restrict interactions to vehicles traveling within approximately $2$--$4$ seconds of one another. This interval is consistent with empirical evidence and microscopic traffic models such as the Intelligent Driver Model, which typically calibrate desired headways in the range of $1.5$--$2.5$ seconds~\cite{Treiber2000, Brackstone1999}. Moreover, road safety guidelines promote the ``two-second rule'' (extended to $3$--$4$ seconds in adverse conditions)~\cite{HighwayCode2022}, further justifying the parameter's value choice.

To illustrate how this parameter affects the congestion score, we provide the following simple example:\\
Consider two vehicles traveling on the same road segment. Let the spatial distance between the vehicles be $d_{i,j}=20\,\text{m}$ and their average speed
$\bar v_{i,j}=10\,\text{m/s}$ ($36\,\text{km/h}$).
The corresponding time headway is
\begin{equation}
h_{i,j}=\frac{d_{i,j}}{\bar v_{i,j}}=\frac{20}{10}=2\,\text{s}.
\end{equation}

Assuming the sensitivity parameter $\gamma=4\,\text{s}$ and time step
$\alpha=10$, the congestion contribution according to Eq.~\ref{eq:score} becomes
\begin{equation}
\text{score}_{i,j}
= \alpha \max\!\left(1-\frac{h_{i,j}}{\gamma},0\right)
= 10 \cdot \left(1-\frac{2}{4}\right)
=5 .
\end{equation}

This positive score reflects a strong leader--follower interaction,
since the vehicles travel with a relatively short temporal separation.
In contrast, if the distance between vehicles increases to
$d_{i,j}=60\,\text{m}$ while maintaining the same speed
$\bar v_{i,j}=10\,\text{m/s}$, the headway becomes
\begin{equation}
h_{i,j}=\frac{60}{10}=6\,\text{s}.
\end{equation}

In this case,
\begin{equation}
\text{score}_{i,j}
= 10 \cdot \max\!\left(1-\frac{6}{4},0\right)=0 ,
\end{equation}
indicating that the vehicles are sufficiently separated and therefore
do not contribute to congestion in the proposed interaction-based metric.

\section{Penalty parameter}\label{app:lambda}
The penalty parameter~$\lambda$ plays a central role in our formulation, since it directly influences both the quality of the solutions and how efficiently solvers perform~\cite{Huang2023, VillarRodriguez2022, Kochenberger2014}. If $\lambda$ is chosen too small, the solver may return infeasible assignments where vehicles are given multiple routes. On the other hand, if $\lambda$ is too large, the penalty term overwhelms the proposed objective, suppressing the intended balance between travel time and congestion.

Different strategies for choosing $\lambda$ have been proposed. In their Mini-scale traffic flow optimization study, Salloum et al.~\cite{Salloum2025} set $\lambda$ equal to the maximum iteration cost of a vehicle–route assignment. This is conceptually close to our approach, although their formulation was restricted to a single route per vehicle.
García et al.~\cite{DiezGarcia2022} introduced three exact approaches to compute valid penalty weights in QUBO: (i) the \emph{sum of absolute coefficients} method, which derives an upper bound on the smallest valid penalty by summing positive and negative contributions in the QUBO polynomial; (ii) the \emph{posiform/negaform} transformation, which rewrites the QUBO into equivalent forms with only non-negative or non-positive coefficients; and (iii) the \emph{Verma--Lewis} method, which derives penalty weights from the largest potential gain obtainable by violating a constraint. All three methods guarantee feasibility preservation, but the Verma--Lewis\cite{Verma2022} method consistently provided penalty weights closest to the best known.

In our formulation, we follow the principle of the Verma-Lewis method.
The formal argument proceeds as follows. Consider any feasible solution and a vehicle $i$ assigned to route $a^\star$, that is,
$x_{i,a^\star}=1$ and $x_{i,a}=0$ for all $a\neq a^\star$. If the selected variable $x_{i,a^\star}$ is flipped from $1$ to $0$, the one-hot constraint for vehicle $i$ becomes violated, since no route is assigned to this vehicle. This violation increases the penalty part of the QUBO objective by $\lambda$.
At the same time, removing $x_{i,a^\star}$ eliminates the linear duration contribution $\pi_{i,a^\star}$ and all congestion interactions involving the selected route $a^\star$.
Since the cost (Eq.~\ref{eq:qubo_cost}) runs over all ordered pairs $i\neq j$, variable $x_{i,a^\star}$ enters both as leader ($w_{i,j,a^\star,a_j}$) and as follower ($w_{j,i,a_j,a^\star}$). The resulting change in the objective function is
\begin{equation}
\Delta \mathcal{Q}
=
+\lambda
-
\pi_{i,a^\star}
-
\sum_{\substack{j\in C\\ j\neq i}}
\sum_{a_j=1}^{k}
\bigl(w_{i,j,a^\star,a_j}+w_{j,i,a_j,a^\star}\bigr) x_{j,a_j}.
\end{equation}

To prevent such a violation from being preferred, the change in the objective must satisfy

\begin{equation}
\Delta \mathcal{Q} > 0\;.
\end{equation}

Substituting the expression for $\Delta \mathcal{Q}$ yields

\begin{equation}
\lambda
>
\pi_{i,a^\star}
+
\sum_{\substack{j\in C\\ j\neq i}}
\sum_{a_j=1}^{k}
\bigl(w_{i,j,a^\star,a_j}+w_{j,i,a_j,a^\star}\bigr) x_{j,a_j}.
\end{equation}

Since the actual congestion contribution depends on the routes selected by the other vehicles, we upper-bound it by the worst-case interaction strength. Using Eq.~\ref{eq:lambda_local} for the selected route $a^\star$, we obtain
\begin{equation}
\Lambda_{i,a^\star}
=
\sum_{\substack{j\in C\\ j\neq i}}
\sum_{a_j=1}^{k}
\bigl(w_{i,j,a^\star,a_j}+w_{j,i,a_j,a^\star}\bigr).
\end{equation}

It is therefore sufficient to require
\begin{equation}
\lambda
\geq
\pi_{i,a^\star}
+
\Lambda_{i,a^\star}.
\label{eq:lambda_condition}
\end{equation}

To guarantee feasibility for all vehicles and route alternatives, the condition must hold for every $(i,a^\star)$. Taking the maximum over all vehicles and route alternatives gives Eq.~\ref{eq:lambda}.

This $1\!\rightarrow\!0$ flip is the critical case for determining $\lambda$, as removing an already selected route produces the largest possible decrease in the objective. All other one-hot modifications either lead to a smaller decrease or increase the objective and therefore do not require a stronger bound.

The connection to Verma--Lewis~\cite{Verma2022} is direct: during the QUBO matrix construction (Algorithm~\ref{alg:qubo_matrix}), each variable $x_{i,a_i}$ is associated with a diagonal contribution $-\lambda+\pi_{i,a_i}$ and off-diagonal congestion interactions. These interactions are represented by the combined directional congestion weights $w_{i,j,a_i,a_j}+w_{j,i,a_j,a_i}$ together with the one-hot cross-terms $\lambda$. Summing the relevant diagonal and off-diagonal contributions associated with $x_{i,a_i}$ yields the bound $\pi_{i,a_i}+\Lambda_{i,a_i}$, which corresponds to the maximum objective reduction obtainable by removing the selected route variable. This confirms that our expression is consistent with the Verma--Lewis approach applied directly to the expanded QUBO matrix.

In practice, $\pi_{i,a_i}$ values are typically one to two orders of magnitude smaller than the corresponding $\Lambda_{i,a_i}$  in our experiments, so the contribution is numerically small. Nonetheless, including it is necessary for the analytical guarantee to hold rigorously.

\section{Construction of QUBO matrix}\label{app:alg_qubo_marix}

Expanding Eq.~\ref{eq:qubo_penalty} for a fixed vehicle $i$ yields

\begin{equation}
\small
\lambda \Big( 1 - \sum_{a_i=1}^{k} x_{i,a_i} \Big)^2
=
\lambda
-\lambda \sum_{a_i=1}^{k} x_{i,a_i}
+ 2\lambda \sum_{1 \le a_i < b_i \le k} x_{i,a_i}x_{i,b_i},
\label{eq:qubo_penalty_expanded}
\end{equation}

where $x_{i,a_i}^2 = x_{i,a_i}$ for binary variables. Consequently, each route variable receives a diagonal contribution $-\lambda$, while each pair of routes of the same vehicle receives a penalty $2\lambda$.

The quadratic term $2\lambda x_{i,a_i}x_{i,b_i}$ is represented by inserting two ordered entries $\mathbf{Q}[q_{i,a_i},q_{i,b_i}]$ and $\mathbf{Q}[q_{i,b_i},q_{i,a_i}]$, each with value $\lambda$. When evaluated by the solver, these contributions sum to the effective coefficient $2\lambda$, ensuring consistency between Eq.~\ref{eq:qubo_penalty_expanded} and Algorithm~\ref{alg:qubo_matrix}. The constant term $\lambda$ is omitted, as it is independent of the decision variables and therefore does not affect the minimization of the objective function.

For congestion interactions, the situation differs. The congestion weights are inherently directional: $w_{i,j,a_i,a_j}$ aggregates all road segments and time steps where vehicle $i$ was the leader over vehicle $j$, while $w_{j,i,a_j,a_i}$ captures the reverse. Because both orderings can occur on different parts of the network and are stored as separate database records, the sum in Eq.~\ref{eq:qubo_cost} runs over all ordered pairs $i\neq j$.
Algorithm~\ref{alg:qubo_matrix} therefore iterates over all ordered pairs and inserts
$\mathbf{Q}[q_{i,a_i},q_{j,a_j}] \mathrel{+}= w_{i,j,a_i,a_j}$.
When the reverse ordering is processed, it additionally inserts
$\mathbf{Q}[q_{j,a_j},q_{i,a_i}] \mathrel{+}= w_{j,i,a_j,a_i}$.

In the QUBO objective function, the binary product is commutative,
$x_{i,a_i}x_{j,a_j}=x_{j,a_j}x_{i,a_i}$.
Therefore, both terms contribute to the same interaction, and the effective coefficient associated with $x_{i,a_i}x_{j,a_j}$ becomes
$w_{i,j,a_i,a_j}+w_{j,i,a_j,a_i}$.
This accounts for both leadership roles without requiring an explicit aggregation step.

\begin{algorithm}[h]
\caption{Construction of QUBO matrix $\mathbf{Q}$ }
\small
\label{alg:qubo_matrix}
\begin{algorithmic}[1]
\State \textbf{Input:} Vehicle set $C$, uniform alternatives $k$, congestion weights $w_{i,j,a_i,a_j}$, duration penalties $\pi_{i,a_i}$

\State \textbf{Map each $(i,a_i)$ to a unique QUBO index $q_{i,a_i}$};
\State Identify the set of vehicles $i$ for which less than $k$ routes was generated and create set of "not-real" route alternatives and set of corresponding indices $\mathcal{N}$.
\State Initialize empty QUBO $\mathbf{Q}\gets\{\}$

\State \textbf{Add congestion weights:}
\For{each ordered pair of vehicles $i\neq j$}
  \For{each $a_i\in\{1,\dots,k\}$}
    \For{each $a_j\in\{1,\dots,k\}$}
      \State $\mathbf{Q}[q_{i,a_i},\,q_{j,a_j}] += w_{i,j,a_i,a_j}$
    \EndFor
  \EndFor
\EndFor

\State \textbf{Compute $\lambda$:}
\State Calculate $\Lambda_{i,a_i}$ via Eq.~\ref{eq:lambda_local}
\State Set $\lambda=\max_{i,a_i}(\pi_{i,a_i}+\Lambda_{i,a_i})$

\State \textbf{Add penalties:}

\For{each vehicle $i \in C$}
  \For{each distinct ordered pair $a_i\neq b_i$}
    \State $\mathbf{Q}[q_{i,a_i},\,q_{i,b_i}] += \lambda$ \Comment{{\footnotesize one-hot off-diagonal}}
  \EndFor
\EndFor

\For{each index $q_{i,a_i}$}
  \If{$q_{i,a_i} \in \mathcal{N}$}
    \State $\mathbf{Q}[q_{i,a_i},\,q_{i,a_i}] += \lambda$ \Comment{{\footnotesize not-real route: discourage selecting}}
  \Else
    \State $\mathbf{Q}[q_{i,a_i},\,q_{i,a_i}] += -\,\lambda \;+\; \pi_{i,a_i}$ \Comment{{\footnotesize real route diagonal}}
  \EndIf
\EndFor

\State \textbf{Output:} QUBO matrix $\mathbf{Q}$, penalty parameter $\lambda$
\end{algorithmic}
\end{algorithm}

\textbf{Numerical example.}
Consider two vehicles $i$ and $j$, each with a single route ($k=1$). On segment $e_1$, vehicle $i$ is ahead of $j$, yielding a directional congestion weight $w_{i,j,1,1}=3.0\,\mathrm{s}$. On a different segment $e_2$, vehicle $j$ is ahead of $i$, yielding $w_{j,i,1,1}=1.5\,\mathrm{s}$.
Algorithm~\ref{alg:qubo_matrix} therefore inserts two ordered entries:
$\mathbf{Q}[q_{i,1},q_{j,1}]=3.0$ and
$\mathbf{Q}[q_{j,1},q_{i,1}]=1.5$.
When the QUBO objective is evaluated, the two entries contribute to the same quadratic term and, due to the commutativity of binary-variable multiplication, are summed, resulting in an effective congestion weight of $4.5\,\mathrm{s}$.

\section{Leiden Clustering}
\label{app:clustering}

To partition the congestion graph into smaller subproblems, we apply the Leiden community detection algorithm~\cite{Traag2019}.\footnote{It is important to note that clustering in this work is not applied in the context of vehicular networking protocols. Instead, it is used purely as a graph community detection technique.}
The choice of Leiden is motivated by both practical and theoretical considerations. Feld et\,al.~\cite{Feld2019} showed that clustering formulated as a QUBO can be unstable, requires dataset-specific penalty tuning, and often fails to respect capacity constraints. In their work on the vehicle routing problem, they instead relied on classical methods such as k-means clustering. Borowski~et\,al.\ \cite{Borowski2020} formalized this principle into a hybrid approach (DBSCAN Solver), where classical decomposition prepares subproblems that quantum annealing can then solve efficiently. These studies highlight that classical preprocessing is an important design choice in hybrid quantum-classical optimization pipelines.

To partition the congestion graph  $G_c = (C,E_c)$ defined in Subsection~\ref{sec:scalability} with weight $w_{i,j}$ as set in Eq.~\ref{eq:cluster_weight}, we apply the Leiden community detection
algorithm. Leiden is an improvement over the Louvain
method and is designed to identify communities that are both modularity
optimal and internally well-connected. The algorithm maximizes the
modularity objective
\begin{equation}
\label{q:leiden}
M_\rho =
\frac{1}{2w_G}
\sum_{i,j}
\left(
w_{i,j} -
\rho \frac{k_i k_j}{2w_G}
\right)
\delta(c_i,c_j),
\end{equation}
where $k_i=\sum_j w_{i,j}$ denotes the weighted degree of node $i$,
$w_G=\frac12 \sum_{i,j} w_{i,j}$ is the total edge weight of the graph, and
$\delta(c_i,c_j)$ equals $1$ when nodes $i$ and $j$ belong to the same
community and $0$ otherwise.
The resolution parameter $\rho$ controls the granularity of the resulting vehicle communities. Lower values produce fewer and larger clusters, while higher values generate a larger number of smaller clusters~\cite{Reichardt2006}.

In this study, we selected $\rho = 4$, which provided a balanced decomposition consisting of multiple clusters that remain strongly connected in terms of congestion interactions. This setting preserves groups of vehicles with high mutual conflicts while keeping individual QUBO instances within solver limits, enabling efficient optimization without fragmenting strongly interacting vehicles across clusters.

The choice of the resolution parameter $\rho$, together with the minimum cluster size $m$ and the maximum number of clusters $L$, influences both the decomposition structure and the resulting solution quality. These parameters were selected empirically to balance cluster cohesion and computational efficiency. While the framework appears stable across the tested instances, a more systematic sensitivity analysis of these clustering parameters remains an important direction for future work.

As with any decomposition-based approach, the selected clustering configuration introduces a trade-off between computational tractability and solution accuracy. This decomposition introduces an approximation because interactions between vehicles belonging to different clusters are not optimized jointly. However, since clustering is constructed directly from the congestion interaction graph, vehicles with the strongest pairwise congestion interactions tend to be placed within the same cluster. To quantify this effect, Subsection~\ref{sec:large} reports the fraction of interaction weight preserved inside clusters, $\eta$. Together with the results in Tables~\ref{tab:qa_large_summary},~\ref{tab:shortest_improvement}, these observations suggest that the decomposition preserves sufficient interaction structure for QAHS to maintain near-optimal global objective values despite the approximation introduced by clustering.\footnote{In our implementation, clustering is an optional step and may be omitted for small-scale instances of TFO.}

The Leiden algorithm proceeds in three phases~\cite{Traag2019}:

\begin{enumerate}
\item \textbf{Local moving phase:} nodes are iteratively moved to
neighboring communities whenever the modularity objective increases.

\item \textbf{Refinement phase:} communities are internally refined to
guarantee that they remain well-connected.

\item \textbf{Aggregation phase:} each community is collapsed into a
super-node, producing a reduced graph on which the procedure is repeated.
\end{enumerate}

In our implementation, clusters smaller than the minimum size are merged with neighboring communities based on the strongest inter-cluster weights. Any remaining clusters are grouped to ensure all subproblems meet the minimum size $m$. This process avoids generating numerous trivial subproblems.

Each cluster $C_\ell \subseteq C$ with $|C_\ell|$ vehicles defines an independent QUBO instance of size $|C_\ell| \cdot k$ variables. These subproblems are solved separately using QAHS and classical solvers. Vehicles outside all clusters (e.g., with negligible interactions) are directly assigned their shortest-duration route. Finally, the local cluster solutions are merged, and total congestion is recomputed across the full city network.

Without clustering, the QUBO formulation requires $O(n^2 k^2)$ pairwise terms, since every vehicle--route pair may interact with every other.
By partitioning the vehicle set into $L$ clusters $\{C_1,\dots,C_L\}$, the effective complexity is reduced to
\begin{equation}
O\!\Big(\sum_{\ell=1}^L |C_\ell|^2 k^2\Big),
\end{equation}
which is significantly smaller whenever clusters are well-balanced and $|C_\ell| \ll n$.
For example, with $n=25{,}000$ vehicles and $k$ route alternatives, the unconstrained formulation contains on the order of $n^2 k^2 \approx 6.25 \times 10^8 k^2$ pairwise terms.
If the vehicles are partitioned into $L=25$ balanced clusters of size $|C_\ell|=1{,}000$, then the total complexity reduces to $25 \cdot (1{,}000^2) k^2 = 2.5 \times 10^7 k^2$ terms.
This represents a reduction by a factor of about $25$, i.e., roughly one order of magnitude.
Thus, clustering not only keeps subproblems within solver limits, but also provides near-quadratic reductions in computational effort.

\section{Solvers details}\label{app:solvers}

\textbf{Quantum Annealing.}
D-Wave's Systems provides cloud-based access to quantum annealing hardware optimized for QUBO problems~\cite{DWaveAdvantage}; access is programmatic via the Python Ocean SDK through the Leap service, where users authenticate with a Leap API token~\cite{DWaveOceanSDK}.
In quantum annealing, the problem is encoded as the ground state of an Ising Hamiltonian and the quantum system evolves adiabatically from an easily prepared initial state toward the target Hamiltonian, ideally remaining in the ground state. The Advantage QPU employs the Pegasus topology, with approximately 5,640 qubits and up to 15 connections per qubit, offering significantly enhanced connectivity over the older Chimera topology ($\sim$6 connections per qubit)~\cite{Boothby2020}.  This increased connectivity allows fully connected problem embeddings of up to $\sim$177 logical variables~\cite{McGeoch2021, McGeoch2014}, but the actual count is problem dependent. Connectivity constraints in the hardware require minor-embedding, which is itself NP-hard and performed classically before annealing~\cite{Choi2008}.
 D-Wave's cloud platform also provides hybrid solvers for larger problems. The \texttt{LeapHybridBQMSampler} handles unconstrained binary quadratic models by combining the quantum annealer with classical heuristics (embedding penalty terms directly into the objective). The \texttt{LeapHybridCQMSampler} supports constrained quadratic models with one-hot constraints enforced directly, but we did not use the CQM solver due to its significantly longer runtime.~\cite{dwave-hybrid} \\

\noindent
\textbf{Gurobi.}
Gurobi is a commercial solver for mixed-integer programming~\cite{gurobi}. It is based on a branch-and-bound framework enhanced with linear programming relaxations and cutting-plane techniques~\cite{Land2010}. QUBO problems can be solved by reformulating them as mixed-integer linear programs (MILPs), introducing auxiliary variables to represent quadratic terms. Although this transformation increases the problem size, it preserves the original QUBO structure, including the one-hot constraints and objective function, allowing for direct comparison with other solvers.
For our simulations, we use Gurobi via the Python \texttt{gurobipy} package distributed on PyPI~\cite{gurobi-pypi}, running under a free academic license through the Gurobi Web License Service~\cite{gurobi-wls}. A per-instance solver execution time limit is set to our quantum annealing solver runtime dynamically. \\

\noindent
\textbf{CBC.}
The COIN-OR Branch-and-Cut (CBC) solver~\cite{CBC} is an open-source MILP solver built on a branch-and-bound framework enhanced by cutting planes. Because CBC does not natively support quadratic terms, we manually linearize the QUBO by introducing auxiliary binaries (similarly to our Gurobi setup).
We accessed CBC through the \texttt{PuLP} Python library~\cite{PuLP} and executed it locally\footnote{Used workstation was equipped with an AMD Ryzen~9~7900X 12-core processor (4.7\,GHz), 64\,GB of RAM,  and 932\,GB of SSD storage}. In practice, however, CBC struggled with medium- or large-scale instances, and therefore we used it mainly as a license-free baseline for small-scale exact optimization.\\

\noindent
\textbf{Simulated Annealing (SA).}
This classical stochastic method provides a baseline for QUBO solving. Inspired by the physical cooling process of materials, the algorithm searches the solution space by applying local modifications: while improvements are always taken, non-improving moves can also be accepted with a probability that decreases as the system "cools". This mechanism helps the algorithm escape local minima and approximate global optima \cite{Kirkpatrick1983, Alnowibet2022}.
We implemented SA using D-Wave's open-source \texttt{neal} library, where the QUBO matrix is converted into a binary quadratic model and sampled with default parameters \cite{Neal}. Due to the lack of optimality guarantees and relatively longer runtimes, SA is more appropriate as a lightweight classical baseline than as a scalable solver. \\

\noindent
\textbf{Tabu Search (Tabu).}
This metaheuristic enhances local search with short-term memory to avoid cycling. Rather than repeatedly revisiting recent solutions, it maintains a dynamic "tabu list" of forbidden moves for a specified tenure, meaning that a move remains disallowed for a fixed number of iterations before it can be reconsidered. This mechanism helps steer the search toward unexplored neighborhoods and balances intensification around promising solutions with diversification into new regions~\cite{Glover1997, Sakabe2022}.
For our testing, we used D-Wave's \texttt{dwave-tabu} sampler~\cite{DWaveTabu2025, Palubeckis2004} with default parameters, which applies tabu-based local search directly to binary quadratic models. Tabu cannot guarantee optimality, it is lightweight, runs locally, and often produces strong heuristic solutions for QUBO problems. Similar to SA, it has longer execution times and is better suited for small- to medium-scale instances.

\section{Description of Overall Workflow}\label{app:workflow}

\begin{enumerate}
    \item \textbf{City map generation.} The road network is extracted from OpenStreetMap, either by specifying a city name or by also providing the center coordinates and a radius to restrict the network to a specific subgraph.
    This flexibility enables simulation of diverse scenarios, from entire metropolitan areas to localized regions of interest.
    Experiments were conducted on multiple cities (Košice, Prague, Cardiff) and their subnetworks.

    \item \textbf{Vehicles generation.} Vehicles are generated with origin–destination pairs either randomly distributed or directed toward the attraction point $C_{\text{att}}$.
    The number of vehicles is set via a parameter $n$, and origin–destination pairs are further constrained by user-defined minimum and maximum lengths ($L_{\min}, L_{\max}$), computed as geodesic distances between the selected points.\footnote{We used this simplification because calculating the exact road/network distance between origin and destination for a large number of vehicles was computationally prohibitive.}
    \item \textbf{Vehicle route alternatives.} For each vehicle, $k$ alternative routes are calculated using the open-source Valhalla routing engine~\cite{valhalla}. The requests to Valhalla are executed asynchronously in batches and processed in parallel.
    To make the routes usable for congestion modeling, they are sampled every $\alpha$ seconds to create a sequence of route points.
    Each point records the vehicle's position, speed, direction, and the closest road segment, so that interactions between vehicles can be tracked in space and time.
    For every route, we also store total distance and travel time.

    \item \textbf{Congestion modeling.} At each time step, vehicles traveling on the same road segment and in the same direction are grouped, and leader–follower pairs are identified.
    For every such pair, the distance between vehicles is computed using the Haversine formula, which accounts for the Earth's curvature.
    This distance is then normalized by their average speed and scaled by a sensitivity factor $\gamma$, producing a congestion score (Eq.~\ref{eq:score}).
    Scores are accumulated across all edges and time steps within the simulation window $w$ to obtain per-edge congestion entries (Eq.~\ref{eq:congestion_entry}), which are later aggregated into pairwise congestion weights $w_{i,j,a_i,a_j}$ (Eq.~\ref{eq:pairwise_weight}) used in the QUBO formulation.

    \item \textbf{Clustering (optional).} For large instances, a congestion graph is constructed and partitioned using the Leiden algorithm ($\rho$ - cluster resolution) to ensure that subproblems remain within solver limits.
    Vehicles outside clusters are assigned their shortest-duration route.
    In rare cases where a solver returns an invalid assignment (e.g., no valid one-hot selection across $k$ routes), the vehicle is also assigned its shortest route.
    Although such situations are uncommon, this fallback ensures that every vehicle receives a valid route and the workflow remains consistent.

    \item \textbf{Optimization (QUBO solvers).} For each cluster (or the full instance when clustering is not applied), a QUBO formulation is constructed and solved using a range of approaches.
    Depending on the problem size (see Section~\ref{sec:instances}), this includes direct execution on the D-Wave's QPU, the D-Wave's QAHS  and classical methods: Gurobi, CBC, Simulated Annealing, and Tabu Search.
    The motivation and setup of each solver are described in detail in Appendix~\ref{app:solvers}.

    \item \textbf{Evaluation and baselines.} Optimized assignments are merged, and total congestion is recomputed across the full network to ensure comparability between clustered and non-clustered vehicles.
    Results are then benchmarked against baselines: shortest-duration and random-route assignments.
    For each solver and baseline, we calculate congestion cost (Eq.~\ref{eq:cong_cost}), solver computational duration, and assignment validity (see Subsection~\ref{sec:performance_metrics}).
    All metrics and solver outputs are written to the MariaDB database.
    
    \item \textbf{Analysis and visualization.} Heatmaps and summary tables are produced to illustrate both city-wide congestion patterns and solver-specific outcomes.
\end{enumerate}

\section{Additional Tables}\label{app:tables}

\begin{table}[H]
\centering
\caption{Minimum, maximum, and average Delta Energy between QAHS and Gurobi for Cardiff and Košice. Vehicles $n$ are rounded up.}
\label{tab:city_delta_energy_compact}
\footnotesize
\setlength{\tabcolsep}{2pt}

\begin{tabular}{
>{\centering\arraybackslash}p{42pt}
>{\centering\arraybackslash}p{30pt}
>{\centering\arraybackslash}p{32pt}
>{\centering\arraybackslash}p{30pt}
>{\centering\arraybackslash}p{30pt}
>{\centering\arraybackslash}p{32pt}
>{\centering\arraybackslash}p{30pt}
}
\hline
&
\multicolumn{3}{c}{\textbf{Cardiff, Wales}} &
\multicolumn{3}{c}{\textbf{Košice, Slovakia}} \\
\cline{2-4}\cline{5-7}
\textbf{Vehicles ($n$)} &
\textbf{Min [\%]} & \textbf{Max [\%]} & \textbf{Avg [\%]} &
\textbf{Min [\%]} & \textbf{Max [\%]} & \textbf{Avg [\%]} \\
\hline
100   & 0.00 & 0.67 & 0.18 & 0.00 & 6.54 & 0.49 \\
500   & 0.00 & 0.80 & 0.26 & 0.00 & 1.10 & 0.62 \\
1,000  & 0.19 & 0.61 & 0.47 & 0.27 & 0.97 & 0.70 \\
1,500  & 0.53 & 0.74 & 0.60 & 0.50 & 0.99 & 0.75 \\
2,000  & 0.58 & 0.58 & 0.58 & 0.61 & 0.93 & 0.79 \\
2,500  & 0.58 & 0.72 & 0.64 & 0.72 & 1.00 & 0.81 \\
3,000  & 0.60 & 0.73 & 0.68 & 0.68 & 1.03 & 0.84 \\
3,500  & 0.68 & 0.69 & 0.69 & 0.78 & 0.98 & 0.89 \\
4,000  & 0.73 & 0.73 & 0.73 & 0.74 & 1.21 & 0.96 \\
4,500  & 0.68 & 0.73 & 0.71 & 0.76 & 1.05 & 0.89 \\
5,000  & 0.62 & 0.79 & 0.71 & 0.91 & 1.31 & 1.03 \\
5,500  & 0.78 & 0.78 & 0.78 & 0.80 & 1.58 & 1.06 \\
6,000  & 0.70 & 0.70 & 0.70 & 1.00 & 1.47 & 1.14 \\
6,500  & 0.78 & 0.78 & 0.78 & 0.94 & 0.94 & 0.94 \\
7,000  & 0.75 & 0.75 & 0.75 & 0.96 & 1.42 & 1.19 \\
7,500  & 0.73 & 0.73 & 0.73 & 1.08 & 1.08 & 1.08 \\
8,000  & 0.83 & 0.83 & 0.83 & 0.87 & 1.02 & 0.95 \\
8,500  & 0.75 & 0.75 & 0.75 & 0.93 & 0.93 & 0.93 \\
9,000  & 0.71 & 0.71 & 0.71 & 0.88 & 0.97 & 0.93 \\
\hline
\end{tabular}
\end{table}

\begin{table}[h]
\centering
\caption{Runtime comparison between QAHS and Gurobi for increasing number of vehicles. All durations are given in seconds.}
\label{tab:runtime_comparison}
\footnotesize
\setlength{\tabcolsep}{2pt}

\begin{tabular}{
>{\centering\arraybackslash}p{42pt}
>{\centering\arraybackslash}p{45pt}
>{\centering\arraybackslash}p{45pt}
>{\centering\arraybackslash}p{45pt}
>{\centering\arraybackslash}p{45pt}
}
\hline
&
\multicolumn{2}{c}{\textbf{Model preparation [s]}} &
\multicolumn{2}{c}{\textbf{Solver runtime [s]}} \\
\cline{2-3}\cline{4-5}
\textbf{Vehicles ($n$)} &
\textbf{QAHS} & \textbf{Gurobi} &
\textbf{QAHS} & \textbf{Gurobi} \\
\hline
100     & 15.07   & 0.87   & 2.99   & 0.62   \\
500     & 29.15   & 3.25   & 3.19   & 1.40   \\
1,000   & 42.17   & 9.68   & 4.80   & 3.10   \\
1,500   & 69.04   & 20.06  & 7.30   & 3.93   \\
2,000   & 93.16   & 34.05  & 9.58   & 6.26   \\
2,500   & 131.28  & 53.81  & 14.01  & 9.59   \\
3,000   & 176.54  & 75.93  & 18.31  & 15.37  \\
3,500   & 252.98  & 111.99 & 24.29  & 23.00  \\
4,000   & 320.37  & 142.98 & 28.77  & 25.37  \\
4,500   & 421.54  & 200.09 & 35.31  & 35.37  \\
5,000   & 476.25  & 206.91 & 39.02  & 32.49  \\
5,500   & 588.42  & 257.98 & 46.36  & 38.57  \\
6,000   & 709.53  & 313.61 & 53.29  & 43.15  \\
6,500   & 809.84  & 382.76 & 63.26  & 63.34  \\
7,000   & 944.36  & 409.87 & 70.23  & 46.67  \\
7,500   & 1082.75 & 500.24 & 80.29  & 80.41  \\
8,000   & 1214.80 & 566.97 & 84.33  & 84.46  \\
8,500   & 1349.13 & 633.11 & 95.23  & 95.39  \\
9,000   & 1517.65 & 717.93 & 100.38 & 100.51 \\
9,500   & 1743.30 & 811.80 & 111.06 & 111.18 \\
10,000  & 1841.23 & 860.79 & 116.80 & 116.94 \\
\hline
\end{tabular}
\end{table}

\section*{Acknowledgment}
During the preparation of this work, the authors used an AI tool to improve language and readability. The ideas and content remain the sole responsibility of the authors.\\

The complete implementation used in this study is available in a public GitHub repository: \url{https://github.com/rusnakrenata/qa_mtc}.

\newpage

\bibliographystyle{IEEEtran}
\bibliography{bibliography}

\begin{IEEEbiography}[{\includegraphics[width=1in,height=1.25in,clip,keepaspectratio]{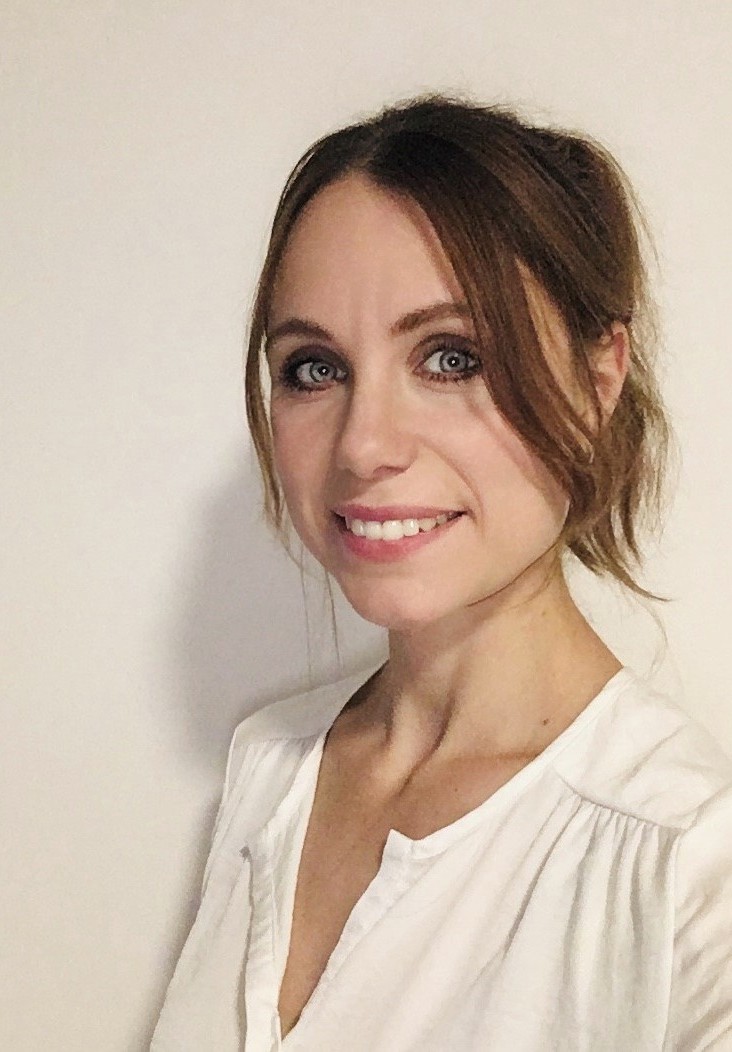}}] {Renáta Rusnáková} received the M.S. degree in mathematical methods of information security from Charles University, Prague, Czech Republic, in 2012. Her master's thesis focused on computational complexity and reduction among NP-hard problems.

She is currently a Ph.D. candidate at the Technical University of Košice (TUKE), Slovakia, specializing in quantum computing and its applications in optimization problems. Her research interests include quantum annealing, combinatorial optimization, and quantum-inspired algorithms for mobility and cybersecurity applications.

Mgr. Rusnáková is actively involved in quantum computing research, contributing to various projects and publications in the field.
\end{IEEEbiography}

\begin{IEEEbiography}[{\includegraphics[width=1in,height=1.25in,clip,keepaspectratio]{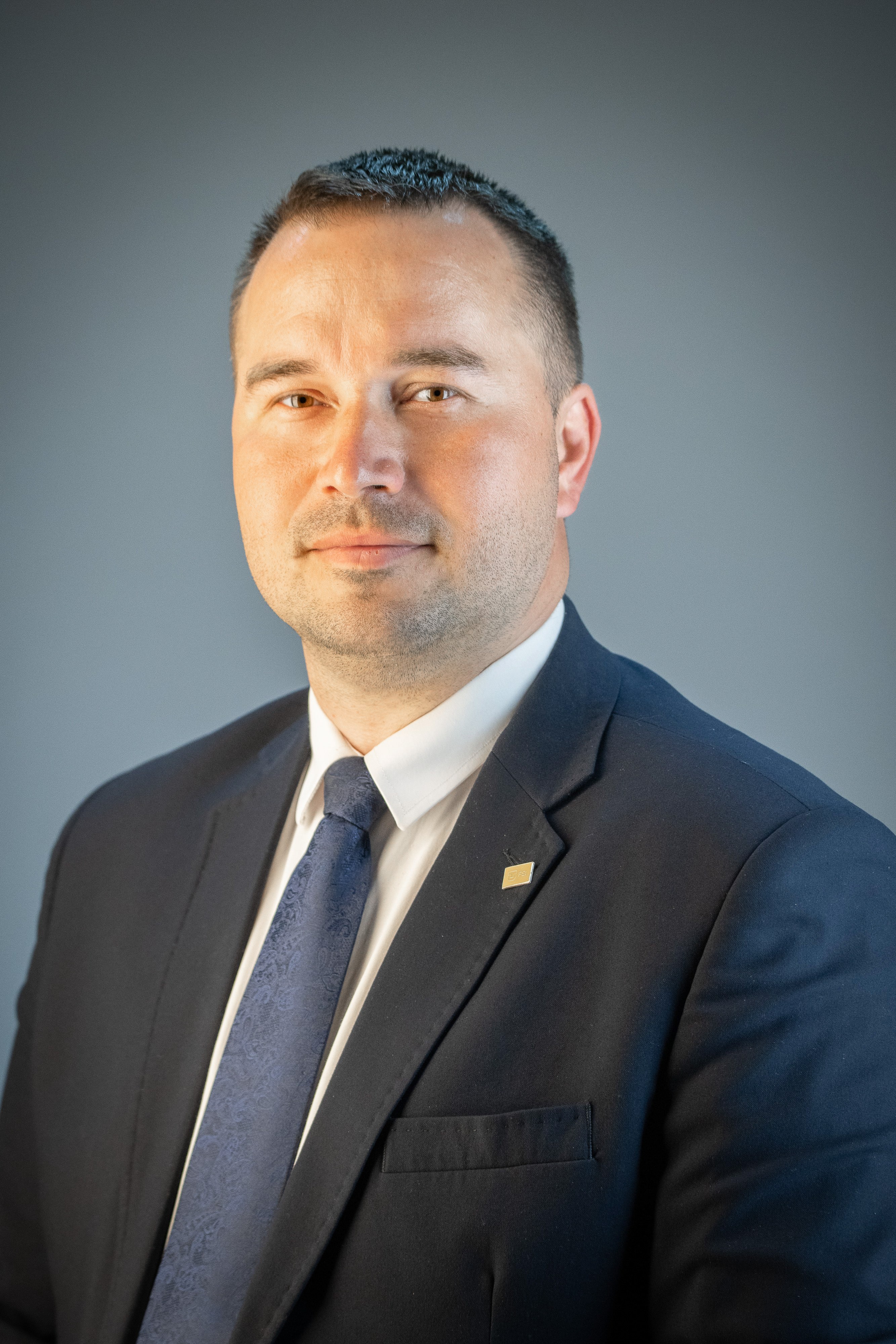}}] {Martin Chovanec} is a university lecturer and researcher in the field of information and cyber security at the Technical University of Košice, specifically at the Department of Computers and Informatics, Faculty of Electrical Engineering and Informatics. His scientific research focuses on cyber security, with an emphasis on intrusion detection systems and network traffic analysis. His expertise includes the design of security architectures, risk management, and data protection in modern IT environments. 

He is the author of several scientific publications contributing to the development of discourse in the field of information system security. 

Assoc. Prof.  Martin Chovanec, PhD. also works as the director of the Institute of Computer Technology at TUKE, where he contributes to the development of the university’s IT infrastructure and the implementation of modern technologies in the management and security of information systems.
\end{IEEEbiography}

\begin{IEEEbiography}[{\includegraphics[width=1in,height=1.25in,clip,keepaspectratio]{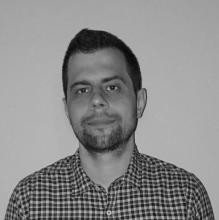}}]{Juraj Gazda} is currently a Vice-Rector for Innovation and Technology Transfer and Professor with the Faculty of Electrical Engineering at the Technical University of Košice (TUKE), Slovakia. 

He has been a guest researcher at Ramon Llull University, Barcelona, and the Technical University of Hamburg-Harburg, and has been involved in development projects for Nokia Siemens Networks (NSN) and Ericsson. In 2017, he was recognized as the Best Young Scientist at TUKE.

Prof. Ing. Juraj Gazda PhD. currently serves on the executive board of IT Valley, an innovation ecosystem supporting collaboration between academia, industry, and public sector, and on the executive board of AI4Slovakia, a national initiative driving the adoption and strategic development of artificial intelligence in Slovakia.
He is also an editor of the KSII Transactions on Internet and Information Systems and a guest editor for Wireless Communications and Mobile Computing (Wiley).
\end{IEEEbiography}

\EOD

\end{document}